\documentclass[prb,reprint,superscriptaddress,longbibliography]{revtex4-2}
\usepackage[english]{babel}
\usepackage{lipsum}  
\usepackage{amsmath}
\usepackage{graphicx}
\usepackage{svg}
\usepackage{tabularx,multirow}
\usepackage{longtable,threeparttablex}
\graphicspath{{figs/}}
\addto\captionsenglish{}
\usepackage{hyperref}
\newcommand{\angstrom}{\text{\normalfont\AA}}
\hypersetup{
 bookmarks=true,		
 unicode=false,			
 pdftoolbar=true,		
 pdffitwindow=false,		
 pdfstartview={FitH},		
 pdfcreator={pdflatex},		
 pdfnewwindow=true,		
 colorlinks=true,		
 linktoc=page,			
 linkcolor=blue,		
 citecolor=blue,		
 filecolor=blue,		
 urlcolor=blue			
}

\begin{document}

\title{Native defects and impurities in talcum quasi-2D layers}
\date{\today}

\author{Gell\'{e}rt Dolecsek}
\affiliation{Department of Physics of Complex Systems, Eötvös Loránd University, Egyetem tér 1-3, H-1053 Budapest, Hungary}
\affiliation{MTA–ELTE Lend\"{u}let "Momentum" NewQubit Research Group, Pázmány Péter, Sétány 1/A, 1117 Budapest, Hungary}

\author{Oscar Bulancea Lindvall}
\affiliation{Department of Physics of Complex Systems, Eötvös Loránd University, Egyetem tér 1-3, H-1053 Budapest, Hungary}
\affiliation{MTA–ELTE Lend\"{u}let "Momentum" NewQubit Research Group, Pázmány Péter, Sétány 1/A, 1117 Budapest, Hungary}
\affiliation{Department of Physics, Chemistry and Biology, Link\"oping University, SE-581 83 Link\"oping, Sweden}

\author{Joel Davidsson}
\affiliation{Department of Physics, Chemistry and Biology, Link\"oping University, SE-581 83 Link\"oping, Sweden}

\author{Viktor Ivády}
\email{ivady.viktor@ttk.elte.hu}
\affiliation{Department of Physics of Complex Systems, Eötvös Loránd University, Egyetem tér 1-3, H-1053 Budapest, Hungary}
\affiliation{MTA–ELTE Lend\"{u}let "Momentum" NewQubit Research Group, Pázmány Péter, Sétány 1/A, 1117 Budapest, Hungary}

\begin{abstract}
Layered semiconductors have recently emerged as capable host materials for novel quantum applications ranging from phonics to sensing. Most studies have focused on artificial layered materials, while natural layered materials, such as talc and other silicates, have remained largely unexplored despite their desirable properties, e.g, wide direct bandgap, low concentration of optically active defects, and low abundance of nuclear spins. In this article, we carry out a comprehensive computational study of pristine and defective talc layers and discuss their potential applications. After investigating bulk properties, such as lattice parameters, band structure, and dielectric constant, we study the electronic structure, charge states, spin and optical properties of vacancy defects, metal, metalloid, and non-metallic impurities. Our results establish the basis for identifying  color centers, electron paramagnetic resonance centers, potential spin quantum bits, and p and n-type dopants.  These findings mature the theory of talc and point toward potential applications in quantum technologies.
\end{abstract}

\maketitle

\section{Introduction}

Today, we are witnessing a revolution in many key areas of science and technology, including but not limited to computing, information processing, and sensing. Nano-scale solid-state structures whose properties are dictated by the laws of quantum mechanics are at the forefront of these developments. As such, van der Waals heterostructures\cite{geim_van_2013} have become widespread and provide a versatile material platform for numerous applications.\cite{liu_van_2016,liang_van_2020,healey_quantum_2022} The most common components of such composite devices are graphene, hexagonal boron nitride, and transitional metal dichalcogenides.

Silicates are common natural earth materials widely used by industry. Among silicates, there are well-known layered materials, such as kyanite, muscovite, biotite, lepidolite, phlogopite, and talcum, all of which have already been mechanically exfoliated\cite{frisenda_naturally_2020} opening the way toward incorporating them into van der Waals heterostructures. Talcum, also known as talc, magnesium silicate hydroxide, and baby powder, has been utilized in many applications\cite{claverie_synthetic_2018,vasic_two-dimensional_2021} including substrate and insulator layer in the van der Waals structures.\cite{mania_spontaneous_2017,barcelos_infrared_2018,gadelha_nano-optical_2021,prando_revealing_2021,nutting_electrical_2021,costa_impacts_2023} Furthermore, talcum has been exfoliated down to a single layer.\cite{alencar_experimental_2015} 

Due to the large direct bandgap, high dielectric constant, and relatively low abundance of paramagnetic isotopes in the lattice, talc is potentially interesting for quantum optics, electronics, and spintronics applications. Despite the desirable properties of talcum layers,  much less attention has been paid to utilizing them as an \emph{active layer} in van der Waals heterostructures. In addition, relatively little is known about structural defects and their effects on the optical, electrical, and magnetic properties of talc. The latest studies\cite{ma_effect_2022,mania_spontaneous_2017,matkovic_iron-rich_2021} focus mostly on abundant contaminants, e.g.\ aluminum and iron. Iron-dopped talc has been proposed as an air-stable magnetic two-dimensional material.\cite{matkovic_iron-rich_2021}

Structural defects play an important role in semiconductors as they can both give rise to new interesting features and degrade desirable properties of the host material. Defect-based spin quantum bits (qubits) and single photon emitters (SPE) in wide-bandgap semiconductors are the building blocks for numerous influential quantum applications.\cite{aharonovich_solid-state_2016,awschalom_quantum_2018} For instance, advances in point defect sensors have opened new horizons in sub-micrometer scale sensing, which is widely used in materials science and biology today.\cite{awschalom_quantum_2018,schirhagl_nitrogen-vacancy_2014} Spin quantum bits in van der Waals materials have become particularly important\cite{gottscholl_initialization_2020,ivady_ab_2020,stern_quantum_2024} by enabling the fabrication of sensing foils\cite{healey_quantum_2022,kumar_magnetic_2022} with high spatial resolution and developing novel van der Waals heterostructures with inbuilt sensing capabilities. This research field is, however, still immature. So far, only a handful of point defect qubits have been identified and utilized in two-dimensional semiconductors, e.g.\ the negatively changed boron vacancy center (VB center) in hBN.\cite{ivady_ab_2020,gottscholl_initialization_2020} The properties of the boron vacancy are suboptimal for high-resolution and high-sensitivity measurements since the VB center has no allowed optical transition\cite{ivady_ab_2020}, couples strongly to its neighbor nuclear spins, and exhibits a short coherence time\cite{haykal_decoherence_2022}. Hence, it is necessary to search for and investigate new point defect qubits in other two-dimensional and quasi-two-dimensional materials.

Defect-based single photon emitters in the wide-bandgap hBN were first reported in 2016\cite{tran_quantum_2016} and an entirely new field was launched demonstrating and developing applicable emitters from the near-infrared to the ultraviolet spectral ranges.  Exfoliated layered semiconductors have important advantages compared to their bulk counterparts, such as high photon collection efficiency due to lack of internal conversion and tunability through strain and electric field.\cite{aharonovich_solid-state_2016,sajid_single-photon_2020} Impurities and defect complexes play an important role since they often create deep energy levels in the band gap of the layered material and give rise to color centers. The large bandgap of talc can accommodate defects and impurity states decoupled from the bands of the host. It is yet to be investigated which defects and impurities can give rise to color centers with desirable properties in talcum layers.

A more traditional use of defects and impurities states in semiconductors is electron and hole doping to enable n and p-type semiconducting materials. Wide-bandgap semiconductors often lack good dopants with thermally accessible shallow defects close to band edges. This limits the application of most wide-band gap semiconductors in high-power electronic and photovoltaics applications. Talcum can host numerous metallic and non-metallic impurities, substituting different atoms of the host. It is an open question of whether these impurities form shallow donor or acceptor states with desirable properties with low formation energy.

In this article, we computationally investigate pristine and defected talcum bulk and quasi-2D layers. First, we present theoretical lattice parameters, band structure, and partial density of the states of bulk talcum obtained by using the HSE06 hybrid functional. The good agreement between theory and experiment establishes the stage for our point defect calculations. In our study, we attempt to cover the most relevant native and impurity-related point defects in talcum layers. We start with vacancies that may be controllably created by high-energy irradiation and ion implantation. Next, we study a large variety of non-metal, metalloid, and metal substitutional impurities in the magnesium and silicon sublattice of talcum. For all defect configurations, we examine the optimized ground state of the atomic configuration, discuss the Kohn-Sham electronic structure, consider possible charge and spin states, calculate hyperfine coupling tensors when it is relevant, and analyze possible optical transitions. Relying on our first principles data, we identify potential single photon emitters in the near-infrared and visible spectral range, potential spin quantum bits with desirable properties, as well as possible p and n-type dopants. Our work establishes the base for advanced control of talc's optical, electrical, and magnetic properties and provides novel defect structures for quantum applications.

\section{Results and discussions}

\subsection{Structure and electronic structure of talc}

\begin{figure}
\includegraphics[width=0.48\textwidth]{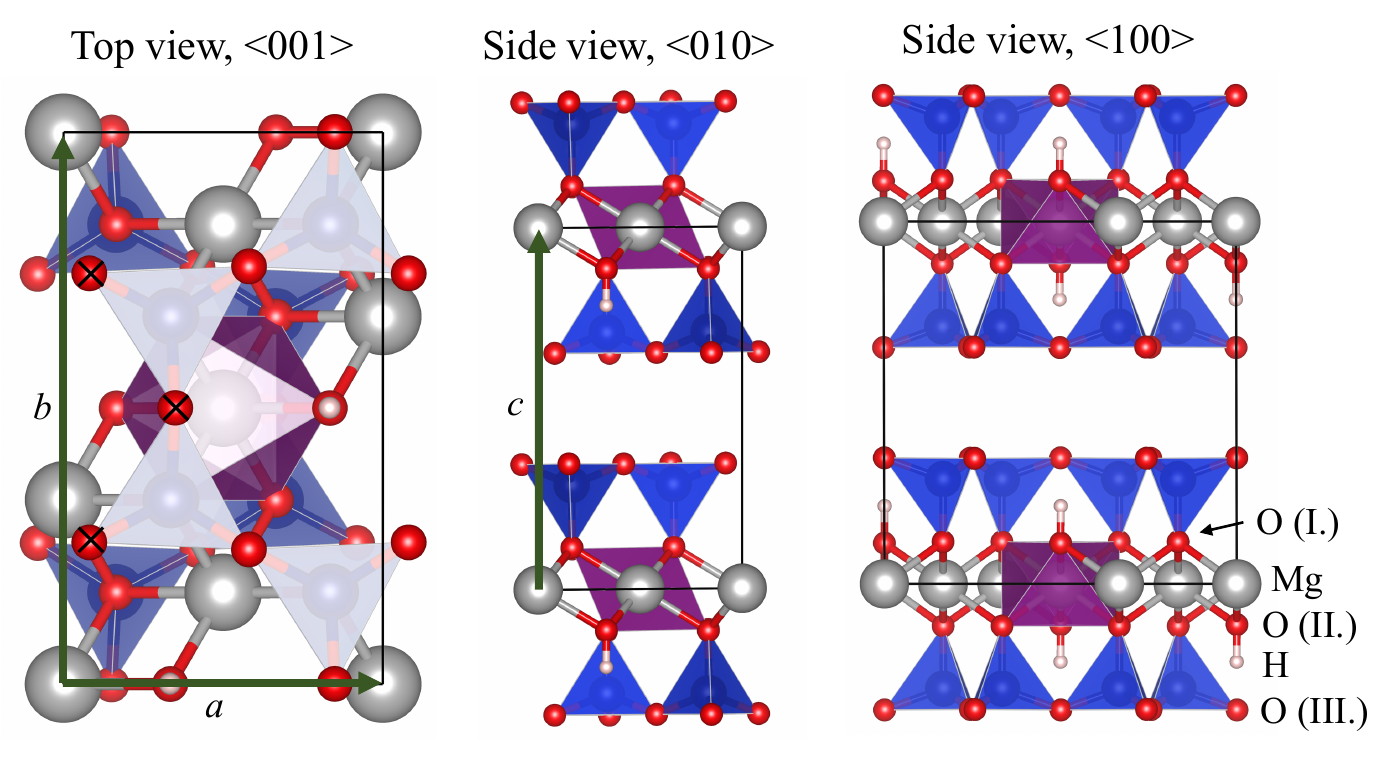}
	\caption{Structure of bulk talc from three different views.  The black solid line depicts the unit cell of talc, while gray, blue, red, and white spheres indicate the position of magnesium, silicon, oxygen, and hydrogen atoms, respectively. The green arrows with labels $a$, $b$, and $c$ show the lattice vectors. The tetrahedra of four-oxygen-coordinated silicon and the octahedra of the six-oxygen-coordinated magnesium are shown by blue and purple polyhedra, respectively. }
	\label{fig:struct}  
\end{figure}

The talc structure as obtained by using HSE06 functional is detailed in Fig.~\ref{fig:struct}. Talc is a layered material with 2.85~\AA\  spacing between the silicate layers. The silicate layers are composed of two layers of SiO$_4$ tetrahedra held together by a layer of magnesium with octahedral coordination. In addition to the four oxygen sitting in the apex of four SiO$_4$ tetrahedra, two OH groups are also connected to a magnesium atom. Each OH group is shared between three magnesium atoms.

\begin{table}[h!]
\caption{Theoretical and experimental lattice parameters of talc.  Experimental parameters Exp.~I.\  and Exp.~II.\  are taken from Ref.~\cite{rayner1973crystal} and Ref.~\cite{ulian2013comparison}, respectively.}
\begin{center}
 \begin{tabular}{c|c|c|c}
 Lattice parameters & Exp. I. & Exp. II. & Calculation \\
 \hline\hline
 a ($\angstrom$)& 5.293 & 5.290 &  5.284 \\
 \hline
  b ($\angstrom$) & 9.179 & 9.173  & 9.131 \\
 \hline
  c ($\angstrom$)& 9.496 & 9.460 & 9.372 \\
 \hline
 $\alpha$ ($^\circ$) & 90.57 & 90.46  & 90.15 \\
 \hline
 $\beta$ ($^\circ$) & 98.91 & 98.68 & 90.47 \\ 
 \hline
 $\gamma$ ($^\circ$) & 90.03 & 90.09 & 89.99 \\
 \hline
\end{tabular}
\end{center}
\label{tab:lat}
\end{table}

The unit cell of talc is triclinic and contains no symmetry operations ($P1$ spacegroup). The calculated lattice parameters, as compared with available experiments, are provided in Table~\ref{tab:lat}. The lengths of the lattice vectors agree very well with the experiment. The largest error of $\approx$1\% is observed in the lattice parameter $c$ specifying the distance and the thickness of the quasi-2D talc layers. The former is dictated by dispersive forces, which are described less accurately in density functional theory. The angles of the lattice parameters show a nearly monoclinic characteristic, i.e.\ $\alpha \approx \gamma \approx 90^{\circ} $ and $\beta \neq 90^{\circ}$. The theoretical results, however, show an orthorhombic line structure, i.e.\  $\alpha \approx \beta \approx \gamma \approx 90^{\circ} $, see Table~\ref{tab:lat}. The largest discrepancy ($\approx$8.3\%)  is observed in the $\beta$ specifying the stacking of the quasi-2D talc layers. This error presumably originates from the same error that causes the underestimation of parameter $c$. Seemingly, the energy landscape over the different stacking configurations of the talc layers is very flat causing issues when trying to find the optimal stacking of the layers. Nevertheless, the inplane parameters $a$, $b$, and $\gamma$ are reproduced with remarkable accuracy and thus we anticipate that the inner structure of the quasi-2D layers is described with high accuracy. For the thickness of the talc single layers, we obtain 6.52~\AA . Finally, we report on a small rotation of the connecting tetrahedra, see the top view in fig.~\ref{fig:struct}. The angle of atoms highlighted by black crosses is 114.9$^\circ$ instead of 120$^\circ$ which corresponds to a perfect alignment. We note that the rotation of the tetrahedra is strain-sensitive and largely depends on in-plane distortions. 

\begin{figure*}
\includegraphics[width=0.9\textwidth]{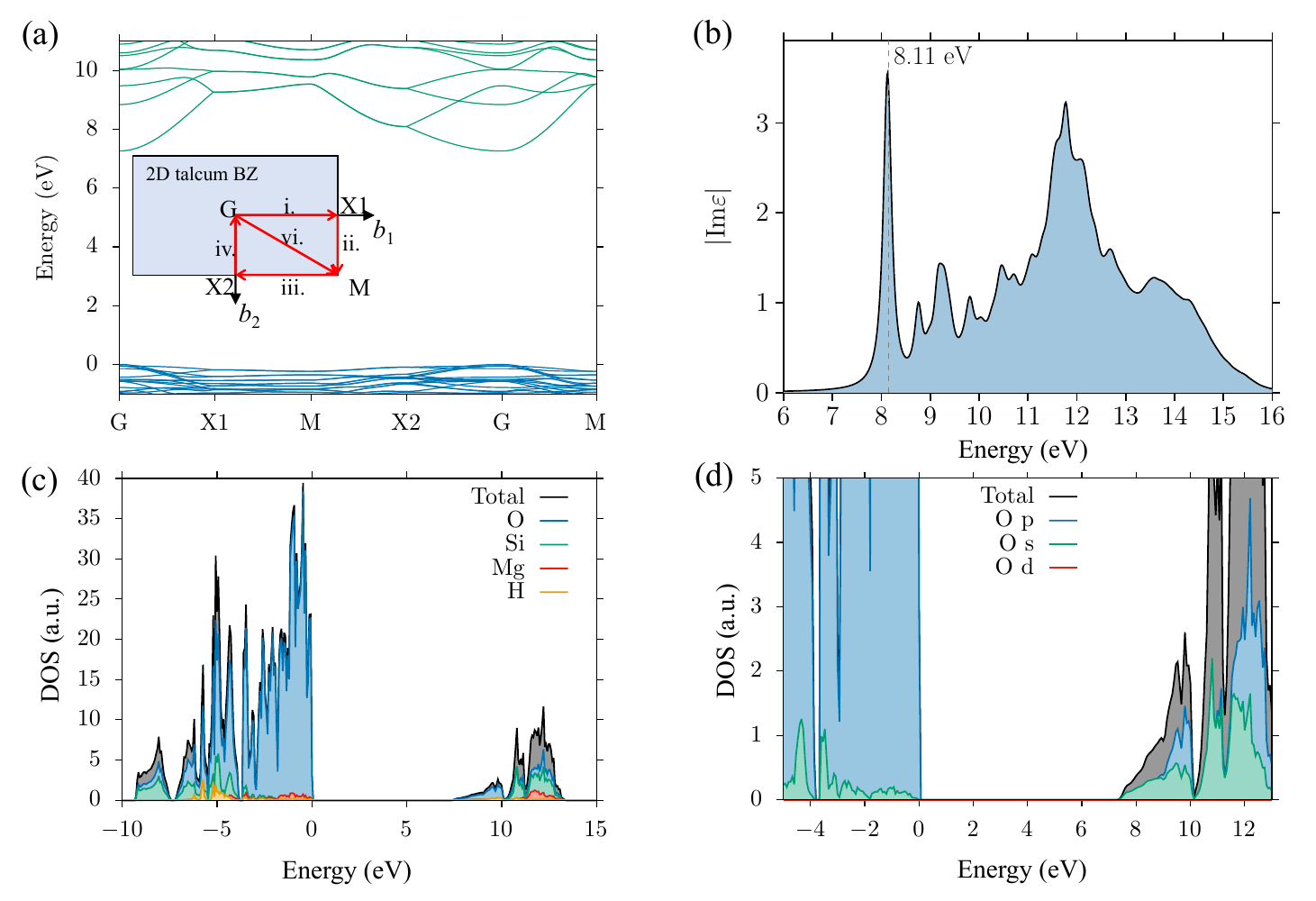}
	\caption{Electronic structure of pristine talc.  (a) Band structure of talc obtained at HSE06 level of theory. The inset depicts Brillouin zone (BZ) with the high-symmetry path used for the band structure calculation. (b) Higher level G$_0$W$_0$+BSE absorption spectrum of talcum. The direct excitonic band-gap of talc is found to be 8.11 eV. (c) and (d) Partial density of states (PDOS) plots of the electronic structure of talc. (c) shows the total DOS and atom-projected PDOS, while (d) displays the total DOS and oxygen orbital-projected PDOS near the band gap.  }
	\label{fig:hse-bandstruct-dos}  
\end{figure*}

\begin{figure}
\includegraphics[width=0.4\textwidth]{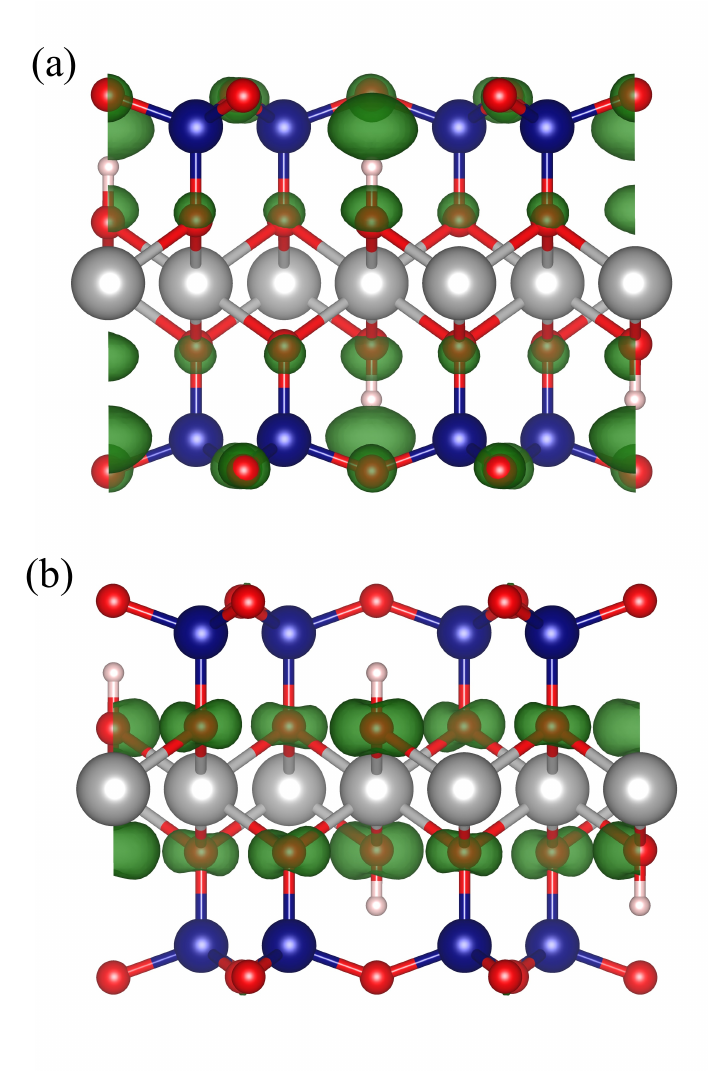}
	\caption{Partial charge densities of the conduction band and valence band edges. a) k-point integrated partial charge density (green lobes) of the lowest energy conduction band state. b) k-point integrated partial charge density of the highest energy valence band.  }
	\label{fig:edge}  
\end{figure}

The computed band structure and the density of states (DOS) of talc are depicted in Figs.~\ref{fig:hse-bandstruct-dos}. The band structure exhibits a 7.27~eV direct bandgap at the gamma point. This value was obtained  from self-consistent field hybrid-DFT calculations with the HSE06 exchange-correlation functional. Previous theoretical calculations reported on 5.3~eV\cite{alencar_experimental_2015} and 5.4~eV\cite{ma_effect_2022} direct band gap using GGA-PBE exchange-correlation energy functional, which tends to underestimate band gaps\cite{ma_effect_2022}. In addition to the Kohn-Sham band structure calculations, we carry out excited state calculations and promote one electron from the valence band edge to the conduction band edge in a 252 atom single-layer supercell model and calculate the band gap from the total energy difference using the $\Delta$SCF method. We employ structural relaxation to enable the emergence of the polaronic effect, which, however, is not observed. The band gap obtained this way is 7.63~eV. Finally, to provide an even better estimate of the band gap, we carry out G$_0$W$_0$+BSE calculation, which is accurate when  polaronic effects are not relevant, see details and the results of the convergence tests in the Supplementary Information. The quasi-particle absorption spectra exhibit an exciton peak at 8.11~eV, see Fig.~\ref{fig:hse-bandstruct-dos}. The reflectivity spectra obtained on the same level of theory can also be found in the Supplementary Information.  Since we are unaware of any experimental reports on the band gap of pristine talcum, thus, our best estimate of the direct excitonic band gap of talc is 8.11~eV.

The valence band edge shows little dispersion, in stark contrast to the conduction band. The conduction band edge is formed by a pair of bands separated from the bulk of the conduction band. Consequently, the DOS near the valence band maximum (VBM) is two orders of magnitude larger than at the conduction band minimum (CBM), see the total DOS in Fig.~\ref{fig:hse-bandstruct-dos}. The atom and orbital decomposed partial DOS reveals that the valence band edge is formed dominantly by oxygen $p$ states, while the conduction band edge is formed by an equal mixture of oxygen $s$ and $p$ orbitals. Both the DOS and the localization of the states at the VBM and CBM are different. As can be seen in Fig.~\ref{fig:edge}, the conduction band edge is mainly localized on the outer O (III.) oxygen atoms of the talcum layer. In contrast, the valence band edge is exclusively  localized on the inner O (I.) and O (II.) atoms. 

Finally, we also calculate the diagonal component of the macroscopic static dielectric tensor by using PBE functional and taking both the electronic and ionic contributions into account. The calculations agree well with available experiments on the permittivity, see Table~\ref{tab:tab_eps}. We note that it has been shown very recently that talc exhibits hyperbolic phonon-polaritons at mid-infrared wavelengths \cite{feres_two-dimensional_2025}.

\begin{table}[h!]
\begin{center}
 \caption{Diagonal elements of the static macroscopic dielectric tensor. The experimental value was obtained from time domain reflectometry immersion method in Ref.~\cite{robinson_measurement_2004}.}
 \begin{tabular}{ c   c   c  |  c }
 \hline
 $\varepsilon_a$ & $\varepsilon_b$ & $\varepsilon_c$ & $\varepsilon_{\text{exp}}$ \\
 \hline
    6.11 & 6.03 & 4.24 & 5.3$\pm0.7$ \\
\hline
\end{tabular}
\label{tab:tab_eps}
\end{center}
\end{table}

\subsection{Formation and defect charge-state transition level energy of vacancy and impurities in talc}

Next, we conduct a comprehensive numerical study of vacancy defects and single substitutional impurities in a single-layer model of talcum on the HSE06 level of theory.  Before we delve into the details of the electronic structure of each defect, we briefly discuss the formation energy of the considered defects from a broader perspective.  To this end, we calculate the formation energy\cite{freysoldt_first-principles_2014} as 
\begin{equation} \label{eq:for}
    E^f[X^q] = E_{\text{tot}}[X^q] - E_{\text{tot}}\text[{bulk}]-\sum_i n_i \mu_i +qE_{\text{F}} + E_{\text{corr}} \text{,}
\end{equation}
where $E_{\text{tot}}[bulk]$ is the total energy of the pristine talc supercell, $E_{\text{tot}}[X^q]$ is the total energy of the defective talc supercell with defect $X$ in charge state $q$, $n_i$ is the number of added and removed atom of index $i$, $\mu_i$ is the chemical potential of atom $i$, $E_{\text{F}}$ is the Fermi energy, $E_{corr}$ is the charge correction term for $q \neq 0$ due to the periodic boundary condition, which is obtained for charged single layer supercells through the jellium charge corrections method introduced in Ref.~\cite{zhang_correcting_2023}. Here, we use $\mu_i$ values of rich phases, therefore, the provided formation energy values are upper limits.

\begin{table}[h!]
\begin{center}
 \caption{Formation energy of neutral defect structures considered in our study. All values are obtained with the HSE06 level of theory in a single-layer talcum model.}
 \begin{tabular}{ c | c }
 \hline
 Defects & E$^f$ (eV)\\[0.5ex]
 \hline
 V$_{\text{H}}$ & 3.340 \\
 V$_{\text{Si}}$ & 16.576 \\
 V$_{\text{Mg}}$ & 10.982 \\
 V$_{\text{O(I.)}}$ & 6.561 \\
 V$_{\text{O(II.)}}$ & 4.520 \\ 
 V$_{\text{O(III.)}}$ & 15.852 \\
 V$_{\text{O}}$V$_{\text{H}}$ & 6.369 \\
 \hline
 Sc$_{\text{Mg}}$ & 1.165 \\
 Zn$_{\text{Mg}}$ & 2.957 \\
 Al$_{\text{Mg}}$ & 2.982 \\
 Mn$_{\text{Mg}}$ & 3.068 \\
 Na$_{\text{Mg}}$ & 5.926 \\
 Ge$_{\text{Mg}}$ & 6.060 \\
 Se$_{\text{Mg}}$ & 10.062 \\
 C$_{\text{Mg}}$ & 10.921 \\
 \hline
 Sc$_{\text{Si}}$ & 3.343 \\
 Al$_{\text{Si}}$ & 3.514 \\
 B$_{\text{Si}}$ & 4.965 \\
  P$_{\text{Si}}$ & 5.680 \\
  Ga$_{\text{Si}}$ & 6.695 \\
 Sb$_{\text{Si}}$ & 6.948 \\
  As$_{\text{Si}}$ & 7.182 \\
 In$_{\text{Si}}$ & 8.259 \\
Bi$_{\text{Si}}$ & 8.350 \\
\hline
\end{tabular}
\label{tab:tab_form}
\end{center}
\end{table}

As shown in Table~\ref{tab:tab_form}, the formation energy of neutral defects and impurities in talc expends over a wide interval, starting at as low formation energies as 1.165~eV of the Sc$_{\text{Mg}}$ defect and ending at as high formation energy as 16.576~eV of the V$_{\text{Si}}$ defect. Among vacancies, the silicon, magnesium, and outer O(III.) oxygens possess especially high formation energies, while inner oxygen vacancies at the O(I.) and O(II.) sites, as well as the OH divacancy, exhibit moderate formation energies. However, hydrogen vacancy possesses relatively low formation energy and is likely to be found in as-grown talc samples. However, high-formation energy vacancies may also be created by electron irradiation, particularly the V$_{\text{O(II.)}}$, V$_{\text{O(I.)}}$, and V$_{\text{O}}$V$_{\text{H}}$. Here, we note that the formation of a single V$_{\text{O(II.)}}$ is rather unlikely compared to the V$_{\text{O}}$V$_{\text{H}}$ divacancy since, for the V$_{\text{O(II.)}}$ defect, a hydrogen atom is left in the lattice, which may diffuse away.

Considering simple substitutional impurities, one can recognize relevant trends in Table~\ref{tab:tab_form}. Most metal substitutions of the magnesium atom possess relatively low formation energy. In particular, scandium, aluminum, zinc, and manganese have low formation energy among the considered impurities. Non-metallic substitutional atoms, such as carbon and selenium, are unlikely to be found at the magnesium site in as-grown talc samples. We also consider both metal and non-metal group~III.\  and group~V.\   impurities at the silicon site. As can be seen, boron and aluminum have low formation energy, while other elements from the same column possess increasing energy with increasing row number. Considering group V.\  impurities at the silicon site, substitutional phosphor and arsenide have moderate formation energy at this lattice site.  

These results indicate that some impurities can incorporate into the lattice of talc with ease. In addition, hydrogen vacancy may , while other vacancies may not be expected in high concentrations in as-grown talcum samples. High-energy electron and neutron irradiation can increase the number of vacancies of different types. Ion implantation may be used to incorporate high-formation energy defects into the lattice.

We note that, the formation energies are calculated for the neutral charge state of the defect.  According to Eq.~(\ref{eq:for}), the neutral charge state provides the maximal formation energy as charged defects may possess lower the formation energy depending on the position of the Fermi energy. Since talc exhibits a large band gap, the Fermi energy can vary in the wide interval, which could lower the formation energies of the charged defects by several electronvolts. This, of course, depends very much on the details of the electronic structure and the possibility of doping talc, which we will discuss in the following sections.

\begin{table}[h!]
\begin{center}
 \caption{Charge transition levels (CTL) of the considered defects and impurities. All values are obtained with the HSE06 level of theory in a single-layer talcum model. The origin of the energy scale is set to be at the valence band maximum of perfect talc. All values are in eV.}
 \begin{tabular}{ c | c | c | c }
 \hline
 Defects & (+/0) & (0/-) & (-/2-)\\  
 \hline
 V$_{\text{H}}$ & --  & 3.01 & -- \\
 V$_{\text{Si}}$ & -- & 2.73 & 3.59 \\
 V$_{\text{Mg}}$ & -- & 0.95 & 2.20 \\
 V$_{\text{O(I.)}}$ & 3.23 & -- & -- \\
 V$_{\text{O(II.)}}$ & 1.40 & -- & -- \\ 
 V$_{\text{O(III.)}}$ & 1.70 & -- & -- \\
 V$_{\text{O}}$V$_{\text{H}}$ & 4.94 & 6.44 & -- \\
 \hline
 Sc$_{\text{Mg}}$ & 6.26 & -- & -- \\
 Zn$_{\text{Mg}}$ & -- & -- & -- \\
 Al$_{\text{Mg}}$ & 7.13 & -- & -- \\
 Mn$_{\text{Mg}}$ & 4.33 & -- & -- \\
 Na$_{\text{Mg}}$ & -- & 0.70 & -- \\
 Ge$_{\text{Mg}}$ & 3.84 & -- & -- \\
 Se$_{\text{Mg}}$ & 4.07 & 4.27 & -- \\
 C$_{\text{Mg}}$ & 3.75 & -- & -- \\
 \hline
 Sc$_{\text{Si}}$ & -- & 0.90 & -- \\
 Al$_{\text{Si}}$ & -- & 0.85 & -- \\
 B$_{\text{Si}}$ & -- & 1.55 & -- \\
  P$_{\text{Si}}$ & 7.12 & -- & -- \\
  Ga$_\textbf{}{\text{Si}}$ & -- & 0.90 & -- \\
 Sb$_{\text{Si}}$ & 3.62 & 3.77 & -- \\
  As$_{\text{Si}}$ & 4.51 & 4.17 & -- \\
 In$_{\text{Si}}$ & -- & 1.26 & -- \\
Bi$_{\text{Si}}$ & 2.00 & -- & -- \\
\hline
\end{tabular}
\label{tab:tab_ctl}
\end{center}
\end{table}

Finally, we calculate relevant charge transition levels for all the considered defects using the following formula\cite{weston_native_2018}
\begin{equation}
    \epsilon(q|q^\prime)=\frac{E_{\text{tot}}[X^q] + E_{\text{corr}}(q) - (E_{\text{tot}}[X^{q^\prime}]+E_{\text{corr}}(q^\prime))}{q^\prime-q} - \epsilon_{\text{VBM}}
\end{equation}
where $E^f(X^q)$ is the formation energy of defect $X$ in charge state $q$ and $\epsilon_{\text{VBM}}$ is the energy of the valence band maximum. Analyzing the position of the charge transition level with respect to the position of the band edge, the valence band maximum in our case, we can identify the stable charge states of the defects at a given value of the Fermi energy. Different Fermi energy values represent different bulk talcum samples with different overall defect concentrations pinning the Fermi energy.

\begin{figure}
\includegraphics[width=0.5\textwidth]{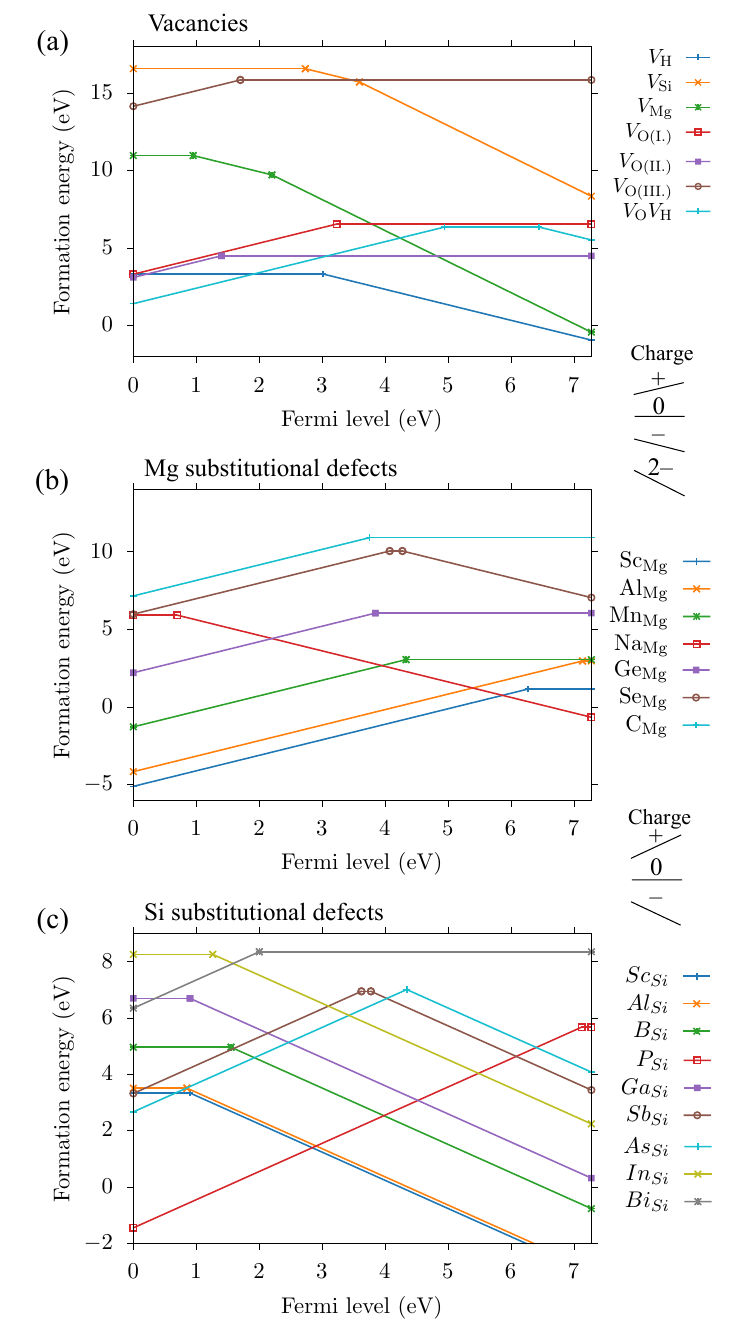}
	\caption{Charge transition level diagrams of talcum point defects. a), b), and c) depict charge and Fermi energy-dependent formation of vacancy-related defects, magnesium substitutional defects, and silicon substitutional defects, respectively.}
	\label{fig:ctl}  
\end{figure}

Table~\ref{tab:tab_ctl}. We discuss the charge transition levels together with the electronic structure of defects in the following subsections. In addition, the content of Tables~\ref{tab:tab_form} and~\ref{tab:tab_ctl} is summarized in the charge transition level diagrams shown in Fig.~\ref{fig:ctl}. These plots help identify general trends, illustrate the relationship between the stable regions of different charge states, and enable comparison of the formation energies of various defects and charge states.

Once the formation energies are defined, we can obtain the thermal equilibrium concentration of each defect in charge state $q$ at temperature $T$ as
\begin{equation}
    C_{X^q}(T) = C_{X\text{-sites}} \, \exp\left(-\frac{E^f[X^q]}{k_B T}\right),
\end{equation}
where $C_{X\text{-sites}}$ is the concentration of sites in pristine talc where defect $X$ can form.

\subsection{Vacancy defects in talc}

\begin{figure*}
\includegraphics[width=0.9\textwidth]{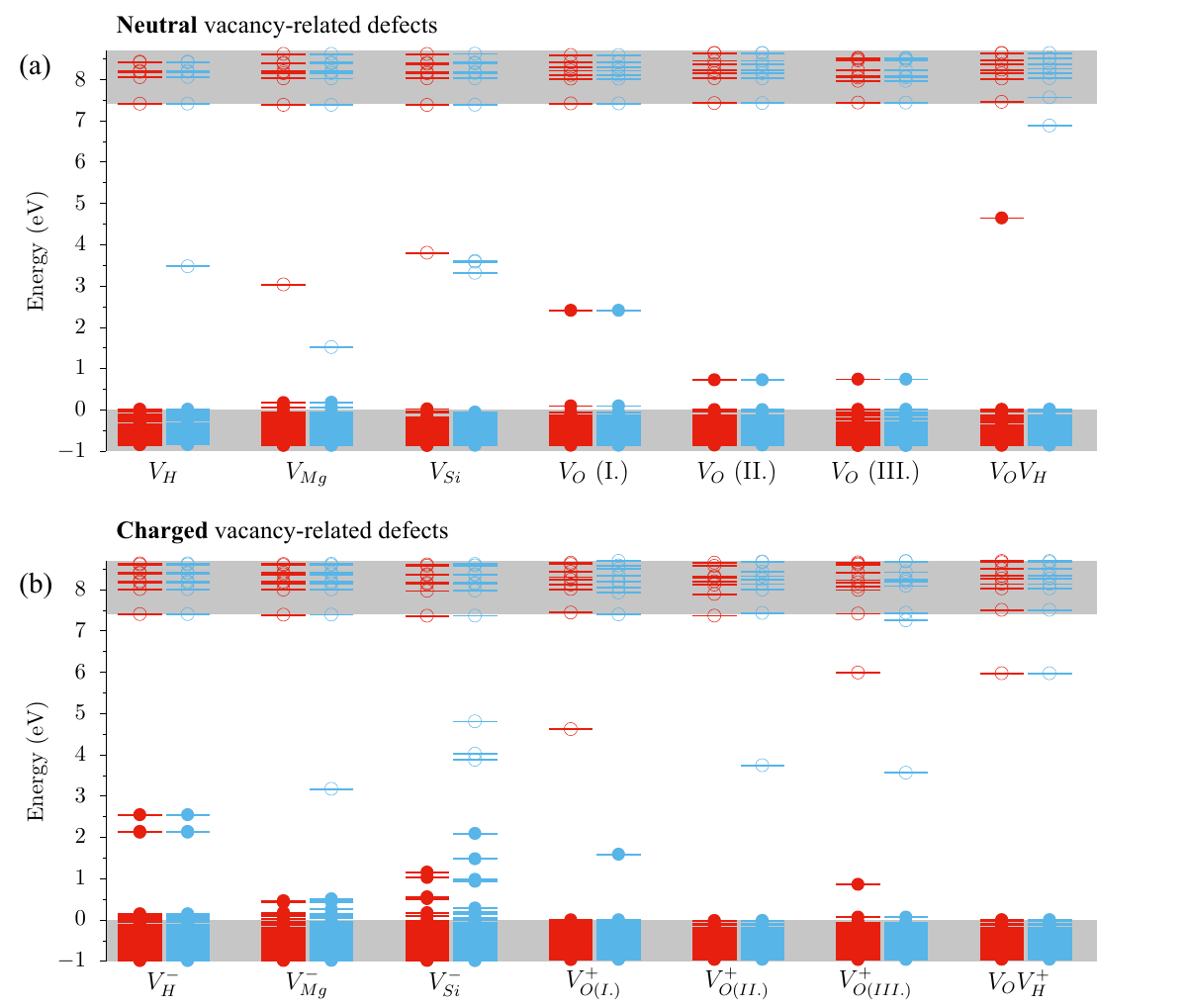}
	\caption{Electronic structure of a) neutral and b) charged vacancy-related point defects in talc. Gray-shaded areas indicate the position of the valence band and the conduction band in a pristine monolayer supercell model, also used to model the defects.  Bars with filled and open circles indicate the position of occupied and unoccupied Kohn-Sham orbitals inside and close to the band gap. Red and blue bars and symbols depict spin-up and spin-down states, respectively. Oxygen sites (I.), (II.), and (III.) are labeled in Fig.~\ref{fig:struct}. $V_OV_H$ represents a divacancy of oxygen (III.) and  hydrogen vacancy next to each other. b) We consider only the most relevant charge state of the defects, i.e.,  the one that is stable toward the middle of the band gap, see Fig.~\ref{fig:ctl}.
    }
	\label{fig:vacancy}  
\end{figure*}

The first class of point defects considered in our study includes vacancies, which can be intentionally created by high-energy irradiation. The Kohn-Sham electronic structures of seven neutral vacancy defects are depicted in Fig.~\ref{fig:vacancy}. We use spin-polarized hybrid DFT, HSE06 functional, and a single-layer model for the calculations. All the investigated vacancy-type defects give rise to defect states inside the band gap. The most relevant deep charge transition levels are also calculated for the considered defects and summarized in Table~\ref{tab:tab_ctl}. 

In addition, in Fig.~\ref{fig:vacancy} we plotted the Kohn-Sham energy level spectrum of charged vacancy defects. In the figure, we consider only the most relevant charge states stable for Fermi-energy values close to the middle of the band gap. In the following, we briefly discuss these defects and highlight their most relevant properties.

\begin{table}[h!]
\begin{center}
 \caption{Table of hyperfine coupling tensors of vacancy-related defects in talc. Only the strongest coupled nuclear spins are listed for each paramagnetic defect. All values are in MHz. }
 \begin{tabular}{ cc | cccccc | c }
 \hline
 defect & atom & $A_{xx}$ & $A_{yy}$ &  $A_{zz}$ & $A_{xy}$  & $A_{xz}$  & $A_{yz}$ & $A_{z}$  \\
 \hline
    V$_{\text{H}}$ & $^{17}$O(II.) &  78.7 & 79.2 &  -319.9 &  0.0  & 13.4  &  -0.5  &  320.2 \\ 
    \hline
    \multirow{ 3}{*}{V$_{\text{Mg}}$} & $^{29}$Si & -85.7  & -85.7 & -85.7 & 0.8 & 0.0 &  0.0  & 85.7 \\
     & $^{17}$O(II.) & -12.5  & 19.4 & -76.0 & -1.8 & 55.5 & 3.2  &  94.1 \\
     & $^{17}$O(II.) & -3.4  & 16.2 & -80.0 & 4.0 & 47.2 & -5.7  &  93.1 \\ 
     \hline
    \multirow{ 2}{*}{V$_{\text{Mg}}^-$} & $^{17}$O(II.) &  -28.4  & 36.7 & -145.7 & 5.8 & 109.6 &  -9.7  & 182.6 \\
    & $^{17}$O(II.) & -28.6 &  37.1  &  -144.9  &  -5.9  &  110.0  & 9.8 & 182.2\\
    \hline
    \multirow{ 3}{*}{V$_{\text{Si}}$} & $^{17}$O(I.) &  -38.0  & -37.6 & 160.8 & 0.0 & 2.3 &  7.3  & 161.0 \\
    & $^{17}$O(III.) & -78.0 &  23.7  &  -10.8  &  -51.6  &  84.6  & 35.2 & 92.3\\
    & $^{17}$O(III.) & 43.8 &  -98.3  &  -8.8  &  -17.5  &  -11.1  & -90.2 & 91.3\\
    & $^{17}$O(III.) & -52.9 &  -5.5  &  -6.3  &  70.9  &  -72.4  & 52.5 & 89.7\\
    \hline
    \multirow{ 3}{*}{V$_{\text{Si}}^-$} & $^{17}$O(I.) & -13.0 & -14.8 & -86.9  & -7.4 & -21.4  & -21.7  & 92.1 \\
 & $^{17}$O(III.)  & -5.0 & -13.5 & -24.2 & 38.5 & -42.5  & 48.4 &  68.8 \\
& $^{17}$O(III.)&  -66.3  & 21.5  & -1.3  &-27.2 &  54.6 &  15.0  & 56.6 \\
& $^{17}$O(III.) &  -7.5 & -27.4 & -39.9  & -9.7 & -11.9 & -29.2  & 50.8 \\
    \hline
    \multirow{ 3}{*}{V$_{\text{Si}}^{-2}$} & $^{17}$O(I.) & 7.8 & 3.2  & -156.2 & 3.4 & -15.1  &  32.9 & 160.3 \\
 & $^{17}$O(III.)  & -110.8  & 32.9 &  10.2 & -42.8 &  73.1  & 19.3  & 76.3 \\
& $^{17}$O(III.) & -10.2 & -13.2 & -26.6 &  10.0 &  -13.2  & 17.1  &  34.3 \\
    \hline
    \multirow{ 1}{*}{V$_{\text{O(I.)}}^{+}$} & $^{29}$Si &  60.5  &  60.5  & 76.3 &  0.0 &  0.0 &  0.0 &   76.3 \\
    \hline
    \multirow{ 1}{*}{V$_{\text{O(II.)}}^{+}$} & $^{2}$H &  32.5  & 32.5  & 32.6  &  0.0  & 0.0  &   0.0  & 32.6\\
    \hline
    \multirow{2}{*}{V$_{\text{O(III.)}}^{+}$} & $^{29}$Si &  28.9  & 36.4 &  29.9  & -0.1  &  0.0 &  -2.9 &  30.1 \\
    & $^{29}$Si & 27.7  & 35.3  & 28.8  &  0.1 &    0.0  & 2.9  & 28.9 \\
    \hline
    \multirow{ 3}{*}{V$_{\text{O}}$V$_{\text{H}}$ } & $^{25}$Mg &  14.6 &  12.6 &  13.3 &  0.0  &  1.2  & 0.0  & 13.4 \\
    & $^{25}$Mg &  12.5  & 13.4  & 12.7  & -0.8 &  -0.6  &  1.0  & 12.8\\
    & $^{25}$Mg &  12.4  & 13.3  & 12.6  &  0.8 &  -0.6 &  -1.0 &  12.7 \\

\hline
\end{tabular}
\label{tab:tab_hyp}
\end{center}
\end{table}

{\it Hydrogen vacancy. }

Due to the removal of the hydrogen atom, the adjacent O (II.) atom relaxes toward the center of the layer by 0.13~\AA\  along the $c$ axis. The defect has a spin-1/2 ground state, and only an unoccupied spin-down KS orbital appears in the band gap in the neutral charge state. The spin density is localized on the first neighbor $^{17}$O atom and gives rise to an $A_{zz} = --319.9$~MHz hyperfine coupling component with the quantization axis $z$ being perpendicular to the plane. Other than oxygen, there are no strongly coupled nuclear spins. The neutral to negative charge transition level is found to be at 3.01~eV above the valence band maximum in the bandgap. Due to the low formation energy of V$_{\text{H}}$, it is expectedly the most likely defect in high purity as-grown talc layers.  The negative charge state has no spin and introduces three fully occupied defects in the lower half of the band gap, see Fig.~\ref{fig:vacancy}. 

{\it Magnesium vacancy. } 

The removal of magnesium from the middle of a talcum layer leads to an outward relaxation of the four O~(I.) atoms of the SiO$_4$ tetrahedra and a substantial inward relaxation of the two OH groups next to the missing magnesium.  Due to the large formation energy of the V$_{\text{Mg}}$ defect and low formation energy substitionals, this defect is not expected in as-grown material and may only appear in irradiated samples.

Interestingly, for the neutral magnesium vacancy, we obtain a spin triplet ground state, which is slightly more favorable than the antiferromagnetic-like open-shell singlet state of the defect ($\approx3$~meV). The small energy difference can be explained by the vanishing exchange energy between the triplet and spin-flipped singlet state due to the involvement of  non-overlapping defect orbitals. Indeed, one of the unoccupied defect states in the band gap is localized on two O~(II.) oxygen atoms, while the other unoccupied state is localized on O~(I.) oxygen on the opposite side of the vacancy.

The neutral to negative charge transition level is found 0.95~eV above the VBM, while the negative to double negative charge transition level is found 2.20~eV above the VBM, see Fig.~\ref{tab:tab_ctl}. The negative and double negative charge states possess spin-1/2 and spin-0 ground states. Since the (0$|$-1) charge transition level is close to the valence band, the magnesium vacancy can presumably be found either in the negative or in the double negative charge state. Interestingly, in both charge states, the valence band edge of talc moves upward due to the appearance of many quasi-localized defect states induced by the negative, double negative V$_{\text{Mg}}$ defect, see Fig.~\ref{fig:vacancy}. Consequently, the bandgap reduces to 6.89~eV and 5.95~eV in the negative and double negative charge states, respectively. In addition to the occupied quasi-localized states at the valence band edge, an unoccupied defect state localized on two nearby O(II.) atoms appears in the bang gap in the negative charge state, Fig.~\ref{fig:vacancy}. Despite the sizable energy gap in the band structure, we do not find a stable excited state geometry for the neutral charge state of the V$_{\text{Mg}}$ defect. The $A_{zz}$ component of the hyperfine tensors of the two O(II.) atoms in the negative charge state is $\sim$182~MHz. Other than these, there are no strongly coupled nuclear spins.

{\it Silicon vacancy. } 

The silicon vacancy in silicates can form various defect structures by reconstructing the oxygen dangling bounds. In talcum, we find a single stable configuration for the silicon vacancy, in which the O(I.) and O(III.) oxygen atoms all relax outward of the silicon vacancy site. At the same time the nearby SiO$_4$ tetrahedra that includes the first neighbor O(III.) oxygen atoms of the silicon vacancy slightly rotates. In the stable configuration, the tetrahedra around the vacancy all rotate in the same direction. We note that the distances of the first neighbor  O(III.) oxygen atoms on the surface are not equal, they are found to be 3.002~\AA ,  3.011~\AA , and 3.033~\AA\ . Rotation of the tetrahedra in the opposite direction may result in slightly different configurations and energy due to the low symmetry of the structure. An  alternative configuration can be imagined by rotating one of the tetrahedra in an opposite direction from the other two tetrahedra, leading to the formation of a long bond between two surface oxygen atoms. We tested this configuration and found it to be unstable and relaxes back to the ground state configuration. 

For the V$_{\text{Si}}$ defect, we obtain a spin-triplet ground state, see the electronic structure in Fig.~\ref{fig:vacancy}. In this high-spin state, the first four neighboring oxygen atoms possess partially occupied dangling bonds that are well separated. We note that the obtained spin state is quite fragile, i.e., an antiferromagnetic-like open-shell singlet state and a spin-2 state lie only 11.3~meV and 26.3~meV above the ground state.  The spin density is localized on the four first neighboring oxygen atoms of the silicon vacancy, see the resulting hyperfine coupling tensors in Table~\ref{tab:tab_hyp}.

Similarly to the V$_{\text{Mg}}$, we do not find a stable optical excited state by promoting one electron from the valence band edge to the defect state. This suggests a continuous nonradiative decay pathway from the excited state to the ground state. A similar mechanism has been reported for other vacancy-type defects, including oxygen in diamond.\cite{thiering_characterization_2016}

Next, we investigate the negative and double negative charge states of the silicon vacancy. The neutral to negative charge transition level can be found at 2.73~eV above the valence band edge, while the negative to double negative charge transition level is positioned at 3.57~eV~eV above the valence band edge. In order, the negative and double negative charge states of the silicon vacancy possess a spin-3/2 and a spin-1 ground states. Similarly, to the neutral charge state, the alternative spin states are close in energy. For instance, we find the energy of the spin-1/2 state 50~meV above the spin-3/2 ground state in the negative charge state.

Due to the addition of extra electron(s), the energy levels are shifted toward the conduction band edge in the negative charge states of the silicon vacancy. As a consequence, numerous occupied and unoccupied defects show up in the bandgap of talcum for V$_{\text{Si}}$(-) defect, see Fig.~\ref{fig:vacancy}, with  numerous excitation possibilities. On the one hand, we find the excitation from the highest occupied defect state to the lowest unoccupied defect state in the minority spin channel unstable, i.e., it relaxes back to the ground state. On the other hand, excitation from the third-highest energy level to the lowest energy unoccupied state gives a stable excited state that lies 0.62~eV above the ground state. Therefore, the negatively charged silicon vacancy fulfills the minimal requirements of a spin qubit and can be a potential candidate for further studies with more adequate techniques, such as CASSCF-NEVPT2 \cite{benedek_accurate_2024}.

Finally, we calculate hyperfine tensors for all possible charge states of the silicon vacancy and provide them in Table~\ref{tab:tab_hyp}.

{\it Oxygen vacancies. }

Due to their relatively low formation energy, oxygen vacancies and V$_{\text{O}}$V$_{\text{H}}$ divacancy may play an important role in irradiated talc samples. According to our findings, oxygen vacancies do not cause major changes to the structure of the talcum layer. For the  O(I.) oxygen vacancy, the first neighbor silicon atom moves vertically toward the inner part of the talcum layer by 0.24~\AA. In the case of the O(II.) vacancy, the hydrogen atom moves closer to the magnesium. As a result, the distance between hydrogen and magnesium reduces to 1.99~\AA from the defect-free distance of 2.66~\AA. Therefore, the O(II.) vacancy can be considered as the complex of a V$_{\text{O}}$V$_{\text{H}}$ divacancy and a H$_{\text{O}}$ substitutional defect. The presence of hydrogen restores the coordination of the magnesium atom, thereby lowering the formation energy compared to the V$_{\text{O}}$V$_{\text{H}}$ divacancy. Despite this alternative interpretation, we consistently refer to the defect as the O(II.) vacancy throughout the paper. For the O(III.) vacancy, we obtain a movement of the first two neighboring silicon atoms of the O(III.) oxygen toward each other. As a consequence, the unperturbed Si-Si distance of 3.047~\AA\  reduces to 2.507~\AA\ .

As can be seen in Fig.~\ref{fig:vacancy}, fully occupied deep defect energy levels appear in the band gap for the three types of single oxygen vacancy defects. Due to the absence of unoccupied defect states in the band gap, the single oxygen vacancy defects do not exhibit optical transitions in the neutral charge state. Furthermore, due to the same reason, oxygen defects can only possess positively charged states. The positive to neutral charge transition levels of the O(I.), O(II.), and O(III.) vacancy can be found at 3.23~eV, 1.40~eV, and 1.70~eV, respectively. Similarly to the neutral charge state, we do not find a suitable lowest energy intra-defect optical transition in the positive charge state, see Fig.~\ref{fig:vacancy}.  On the other hand, the charged oxygen vacancies become spin active as they possess a spin-1/2 ground state. For the O(I.) vacancy, the spin density is localized on the silicon atom next to the vacancy site and gives rise to an $A_{zz}=76.3$~MHz secular hyperfine component for the $^{29}$Si isotope. In the case of the O(II.) vacancy, considerable hyperfine interaction is only observed with the hydrogen atom substituting the oxygen atom with an $A_{zz} = 32.6$~MHz. In the case of O(III.) vacancy, the spin density is localized on two first neighbor silicon atoms. The corresponding hyperfine values are listed in Table~\ref{tab:tab_hyp}.

{\it Oxygen-hydrogen divacancy. } 

Similar to the oxygen vacancies, the oxygen-hydrogen divacancy does not induce major changes in the structure of talcum. Due to the removal of the two atoms, the first neighbor magnesium atoms relax toward the surface slightly. As a consequence, the Mg-Mg distance increases by 0.03~\AA . 
Concerning the electronic structure, a single partially filled deep energy level is found in the band gap for the V$_{\text{O}}$V$_{\text{H}}$ divacancy, see Fig.~\ref{fig:vacancy}. Consequently, the divacancy is spin active in the neutral charge state. The spin density is localized on three first neighbor Mg atoms of the divacancy with a hyperfine splitting of A$_{zz} = 12.8$-$13.5$~MHz, see Table~\ref{tab:tab_hyp}  Due to the absence of occupied and unoccupied defect states in the same spin channel, we can not discuss optical excitation in case of the neutral divacancy.

The divacancy can possess both positive and negative charge states. We find the positive to neutral and the  neutral to negative charge transition levels at 4.94~eV and 6.44~eV above the valence band maximum, respectively.

\subsection{Impurities in talc}

Next, we examine numerous substitutional defects incorporating alkaline, alkaline earth, transition metal, metalloid, and nonmetal contaminants. The studied substitutional defects may be found in natural samples, created by ion implantation, and grown in by using contamination sources.

Interlayer contaminants are also expected in bulk talc samples, however, they may diffuse at room temperature (see a similar study in hBN in Ref.~[\onlinecite{weston_native_2018}]), which makes them less interesting from the application point of view. Therefore, our study focuses on immobile substitutional defects of higher technological relevance.

Solving the Kohn-Sham equation, we observe occupied and unoccupied states in the band gap in most cases, see Figs.~\ref{fig:subs}-\ref{fig:subs}. In the following, we briefly discuss these defects and highlight their most relevant properties.

\subsubsection {Magnesium substitutional defects}

\begin{figure*}
\includegraphics[width=0.9\textwidth]{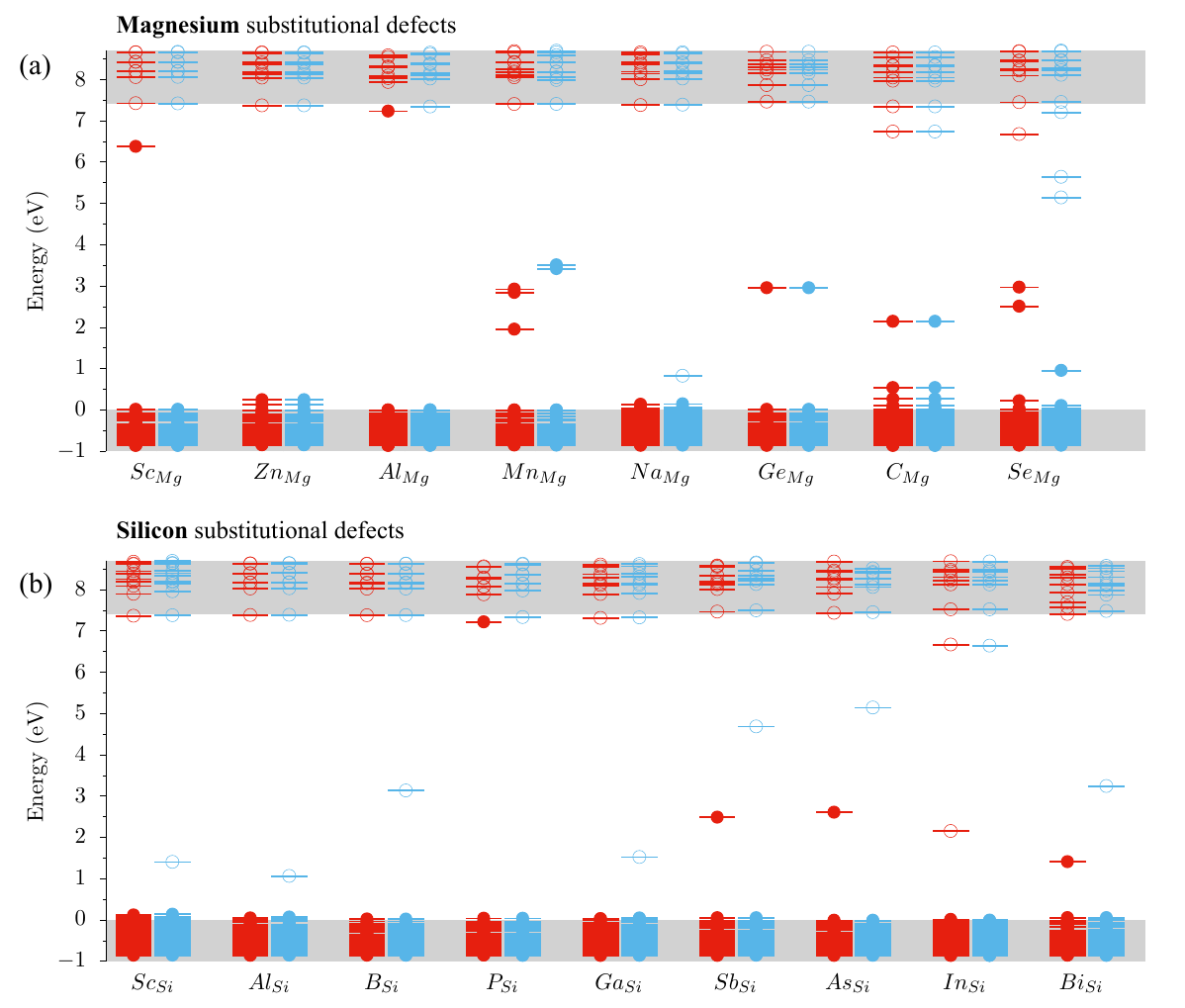}
	\caption{Electronic structure of a) magnesium and b) silicon substitutional defects in talc. Gray-shaded areas indicate the position of the valence band and the conduction band in a pristine monolayer supercell model.  Bars with filled and open circles indicate the position of occupied and unoccupied Kohn-Shame orbitals inside and close to the band gap. Red and blue bars and symbols depict spin-up and spin-down states, respectively. }
	\label{fig:subs}  
\end{figure*}

{\it Scandium impurity}

Sc$_{\text{Mg}}$ has a very low formation energy, see Table~\ref{tab:tab_form}.  The trioctahedral layer of talc incorporating a scandium impurity resembles the pristine talc layer, the Sc-O(II.) distance is increased by only 5\% compared to the Mg-O(II.) distance.
Scandium possesses one extra valence electron compared to magnesium and forms a half-occupied donor state below the conduction band edge, see Fig.~\ref{fig:subs}. The positive to neutral change transition energy level is 1.01~eV below the conduction band edge. Due to the low formation energy of the Sc$_{\text{Mg}}$ defect and the proximity of the positive to neutral charge transition level to the conduction band, the formation energy of positively charged Sc impurity becomes negative in semi-insulating and p-type samples.

{\it Zinc impurity}

The incorporation of zinc into the magnesium layer also requires minimal energy. Considering the structure of the Zn$_{\text{Mg}}$ defect, we observe minor changes. The Zn-O bond lengths are only $\approx$1.5\%  larger than the pristine Mg-O bond lengths. Concerning the electronic structure, we observe 4 fully occupied split-off defect states close to the valence band maximum. Zinc substitutional defect does not possess spin or enable optical transition, however, it lowers the band gap of talc.

{\it Aluminium impurity}

The structure of aluminum substitutional defect exhibits a small contraction of the bond lengths, i.e.\   the Al-O distance reduces by $\approx$8\% . Similar to scandium, aluminum possesses one extra valence electron compared to magnesium. The loosely bonded electron forms a defect state very close to the conduction band edge. The positive to neutral charge transition level is 0.14~eV below the conduction band minimum. Note that the binding energies of donor and acceptor states obtained in our calculations are slightly overestimated due to finite-size effects. Thus the actual donor energy level could be found even closer to the band edge. Aluminium, which is a common impurity of talcum, can serve as a good electron donor and enables n-type doping when located in the magnesium sublattice.

{\it Manganese impurity}

Similar to the impurities mentioned so far, the formation energy of the manganese substitutional defect is also low, see Table.~\ref{tab:tab_form}. The substitutional manganese atom causes minimal changes to the structure, i.e., the Mn-O(I.) distance increases by only 0.5\% . On the other hand,  there are five deep defect states that appear in the two spin channels of the band gap of talc. These states are formed by the split d orbitals of the manganese impurity. Since all states of the defect are occupied, there are no roots for optical transitions. The manganese impurity has a spin-1/2 ground state in the neutral charge state.

The defect can be positively charged and the positive to neutral charge transition level is 4.33~eV above the valence band edge. For the positive charge state, we find a spin-triplet ground state. In addition, the electronic structure of the  positive charge state, which is not shown, enables an optical transition in the minority spin channel. The estimated using PBE functional is 0.72~eV. Therefore, the positively charged manganese substitutional defect, with low formation energy, may serve as a spin qubit. We note that caution is needed for transitional metal impurities due to the use of PBE and hybrid functional\cite{ivady_role_2013,ivady_theoretical_2014}, therefore additional studies may be required to fully understand manganese impurity in talcum.

{\it Sodium impurity}

The sodium substitutional has a formation energy doubled that of the other impurities considered so far, see Table~\ref{tab:tab_form}. This is due to the increased structural distortion induced by the sodium impurity. The O(II.) oxygen atoms and the connected hydrogen atom shift away from the sodium atom. The largest O(I.)-O(II.) distance, i.e.,  the diagonal of the sodium-centered octahedron, increases by 11.\%  from 4.109~\AA\  to 4.588~\AA .

Sodium has one less valence electron compared to magnesium, thus a loosely bonded hole state appears close to the valence band, see Fig.~\ref{fig:subs}. Furthermore, two fully occupied split-off valence band states appear in the band gap, presumably due to the structural distortion caused by the sodium impurity. The neutral to negative charge state is 0.70~eV above the valence band maximum.

{\it Germanium impurity}

The neutral germanium substitutional defect does not significantly alter the structure, the Ge-O distances increase by $\sim$8\%  compared to the Mg-O distances. The formation energy of the defect is similar to the sodium substitutional. The germanium atom has two more valence electrons than the magnesium that form a fully occupied defect state close to the middle of the band gap, see Fig.~\ref{fig:subs}. No other defect states are observed, thus the germanium defect is magnetically and optically inactive. The positive to neutral change transition level is 3.84~eV above the valence band maximum.

{\it Carbon substitunal}

Carbon and selenium impurities have high, larger than 10~eV formation energy. Substitution of magnesium with a nonmetallic carbon atom breaks the trioctahedral structure by shifting out of the center of the octahedra and bonding strongly to the O(II.) oxygen atoms. The C-O(II.) distance reduces by 0.68~\AA\  to 1.39~\AA , explaining the large formation energy obtained for this defect. Due to the distortions and the additional elections, numerous defect states are found in the bandgap of talcum. Three defect states occupied with six electrons can be found close to the valence band edge and an empty defect state close to the conduction band edge. The neutral carbon substitutional defect has a singlet spin state. Since both empty and occupied states are inside the bandgap, optical excitation is possible between the defect states. Therefore, the C$_{\text{Mg}}$ defect may serve as a color center of talc with a zero phonon line (ZPL) energy of 1.92~eV, although the high formation energy of the defect may prevent the formation of a simple substitutional defect. The positive to neutral charge transition level of the carbon vacancy is 3.75~eV above the valence band.

{\it Selenium impurity}

The trioctahedral layer of talc incorporating a nonmetallic selenium atom is largely distorted. However, the displacement of the first neighbor atoms is smaller than that of the carbon substitutional. In the case of selenium atoms, the oxygen atoms  move toward the nonmetallic impurity and the selenium atom remains in the magnesium site, in contrast to carbon impurity. A zoo of selenium-related occupied and unoccupied defect states is found in the band gap. Since there are empty and occupied states, we can charge the defects both negatively and positively. The positive to neutral and the neutral to negative charge transition levels are 4.07~eV and 4.27eV above the valence band maximum. Interestingly, the neutral selenium impurity possesses a spin-1 ground state. Furthermore, optical emission is also possible from this defect with a ZPL energy of $\approx$3.16~eV. These properties suggest that it could serve as a spin qubit in talc. However, due to the narrow Fermi energy region in which the neutral charge state is stable, as well as the high formation energy, the neutral selenium substitutional defect is a less promising candidate.

\subsubsection {Impurities of silicon}

{\it Scandium impurity}

Similar to the magnesium substitutional site, the neutral Sc$_{\text{Si}}$ defect has unexpectedly low formation energy. It is higher than the formation energy of the Sc$_{\text{Mg}}$, however; comparable to the formation energy of Al$_{\text{Si}}$, see Table~\ref{tab:tab_form}, which is an abundant defect in talc. Examining the structure of the Sc$_{\text{Si}}$ defect, we obtain a significant change in the Sc-O(III.) bond lengths. After the relaxation, the tetrahedron centered around the scandium is stretched along the Sc-O(III.) bonds by about 15\% due to increased covalent radius of scandium. To accommodate a larger  tetrahedron, the tetrahedra next to the ScO$_4$ unit are rotated. The electron configuration of scandium possesses one less valence electron compared to silicon, which forms an unoccupied state above the valence edge, see Fig.~\ref{fig:subs}. Sc$_{\text{Si}}$ thus forms an acceptor state. The neutral to negative charge transition level is 0.90~eV above the valence band edge.

{\it Aluminium impurity}

The volume of the aluminum-containing tetrahedron has only slightly changed compared to a pristine SiO$_4$ tetrahedron, i.e., the O(III.)-O(III.) distances increased only by about 5\%. However, the  aluminum atom shifts from the symmetric middle point toward one of the planes of the tetrahedra. One of the Al-O(III.) bond lengths increases by $\approx12\%$ while other Al-O bond lengths by only 5\%. Similarly to the scandium impurities, the neutral Al$_{\text{Si}}$ has a lower formation energy, slightly larger than the formation energy of the neutral Al$_{\text{Mg}}$ defect.
In the band gap, we can find an unoccupied defect state above the valence edge, see Fig.~\ref{fig:subs}. Therefore the Al$_{\text{Si}}$ is a possible acceptor and enables p-type doping of talc. The neutral to negative charge transition level is 0.85~eV above the valence band edge.

{\it Boron impurity}

The boron impurity causes significant changes to the structure of talc. As the boron possesses three valence electrons and its covalent radius is smaller than that of silicon, the tetrahedron around the boron impurity breaks. Instead, a trigonal planar structure forms where the boron atom bonds to two O(III.) oxygens and one O(I.) oxygen. The increased formation energy of the impurity reflects the frustrated structure caused by the boron impurity.  In the band gap, we can find a deep unoccupied defect state well above the valence band edge in contrast to the case of aluminum and scandium. The difference between the defect neutral to negative charge transition level and the valence edge is 1.55~eV.

{\it Phosphorus impurity}

Substitution of silicon with a phosphorus atom causes only a slight change in the tetrahedral SiO$_4$ layer of talc. The O(III.)-O(III.) distances of the phosphorus-centered tetrahedra contracts only by approximately 5\% .  Since the phosphorus impurity possesses one more valence electron compared to silicon, we observe an occupied defect energy level below the conduction band edge, see Fig.~\ref{fig:subs}. Consequently, phosphorus can serve as a good electron donor and enables n-type doping. The positive to neutral charge transition level is 0.15~eV below the conduction band edge.

{\it Gallium impurity}

Similarly to the aluminum substitutions defect, the neutral gallium substitutional atom also breaks the tetrahedral symmetry of the GaO$_4$ unit by the gallium atom shifting toward one of the boundary planes of the tetrahedron. As a consequence, two O(III.) atoms on the surface are pushed apart, which causes tension and rotation of the connecting tetrahedra. This may explain the increase in the formation energy of the defect.  Since gallium has one less valence electron than silicon, an unoccupied defect energy state can be found in the band gap. The negative to neutral charge transition level is 0.90~eV.

{\it Antimony impurity}

The tetrahedral layer of antimony doped talcum is significantly distorted due to the large, 1.96~\AA  Sb-O(III.) bond length, which is $\approx 21\%$ larger than the 1.62~\AA\  Si-O(III.) bound length. The antimony atom is almost coplanar with the three surface O(III.) oxygen atoms in the relaxed structure.  Since the neutral antimony impurity possesses one more valence electron compared to silicon, we can observe one half-occupied defect energy level deep in the band gap, see Fig.~\ref{fig:subs}. 
The positive to neutral and the neutral to negative charge transition levels are 3.62~eV and 3.77~eV above the valence band edge, respectively.

{\it Arsenic impurity}

The electronic structure of a neutral arsenic atom positioned in the metalloid group in the periodic table is very similar to that of the antimony impurity, i.e.,  a half-occupied defect energy level appears deep in the band gap, see Fig.~\ref{fig:subs}. The positive to neutral and the neutral to negative charge transition levels are 4.51~eV and 4.17~eV above the valence band edge exhibiting a negative-U property. On the other hand, the atomic structure shows significant differences. The arsenic atom also destroys the oxygen tetrahedron structure, however; it moves toward the O(III.)-O(I.)-O(III.) layer in the optimized geometry.  O(III.)-As-O(I.) angle decreases from the tetrahedral $109.6^{\circ}$ to $101.8^{\circ}$. To accommodate the distortion and extension of the AsO$_4$ tetrahedra around the As$_{\text{Si}}$ defect significantly, by approximately 15$^{\circ}$ rotates compared to their original orientations.

{\it Indium impurity}

Similar to antimony, the neutral indium substitutional defect induces large distortion to the perfect talc structure by rotating nearby tetrahedra by $\approx 10\%$. This is due to the large, 1.993~\AA\  covalent bond length of the indium and the O(III.) oxygen atoms, which is $\approx 22$\%  larger than the Si-O(III.) bond length of  1.621~\AA .  Indium has one less valence electron than silicon and a resonant d subshell. The electronic structure of the indium substitutional in talc has one deep half-occupied energy level and an empty energy level below the conduction edge. The neutral to negative charge transition level is 1.26~eV above the valence band maximum.

{\it Bismuth impurity}

Finally, we discuss the case of the substitutional bismuth impurity.  The bismuth atom possesses one more valence electron than silicon and it can be found in the metalloid group with a resonant d subshell. The electronic structure of the Bi$_{\text{Si}}$ defect is different from the shallow donor P$_{\text{Si}}$ defect, and it resembles the electronic structure of Sb$_{\text{Si}}$ and As$_{\text{Si}}$. Indeed, we  observe one half-occupied defect energy level in the band gap, see Fig.~\ref{fig:subs}. The positive to neutral charge transition level is at 2.00~eV. In the tetrahedral layer of talc, the Bi-O(III.) distance (2.09~\AA ) is $\approx 29\%$  larger than the Si-O(III.) bond length. Due to the enhanced volume of the BiO$_4$ unit, the surrounding SiO$_4$ rotates the tetrahedra silicate layer of talc by $\approx 30^{\circ}$.

\section{Discussion and summary}

In this article, we examined the properties of pristine and defective talcum, a naturally occurring layered wide band gap semiconductor. In our study, we employed the workhorse of first principles semiconductor studies, the HSE06 functional together with supercell models for defected structures. After calculating the lattice parameters and the band gap, we examined vacancy defects and substitutional impurities in a single-layer supercell model. We computed a wide range of metallic, metalloid, and nonmetallic contaminants either at the silicon or magnesium site. 

Considering the energetics of the defects, we found both low and high formation energies for vacancies. The most common vacancy defect in thermal equilibrium is the neutral hydrogen vacancy. Considering its 3.3~eV formation energy and the relatively low growth temperatures used in hydrothermal synthesis processes (300-600~$^{\circ}$C), the equilibrium concentration of the V$_\text{H}$ defect is expected to be between 10$^{-6}$ and 10$^{3}$~cm$^{-3}$. Even at elevated growth temperature of 1000~$^{\circ}$C, the concentration would only reach approximately 10$^{9}$~cm$^{-3}$. Further vacancy-type defects could be present in even lower concentrations. Since other native defects, such as antisite defects, are not expected, as-grown talcum single crystals can essentially be considered native defect-free.

Considering impurities, we found the formation energy for the neutral scandium to be as low as 1.165~eV, while for Al$_{\text{Mg}}$, Zn$_{\text{Mg}}$, and Mn$_{\text{Mg}}$, it is close to 3.0~eV. Regarding the silicon site, the Sc$_{\text{Si}}$, Al$_{\text{Si}}$, and B$_{\text{Si}}$ defects can be considered low formation energy impurities. This means that these impurities may appear in talcum in high concentration, depending on the availability of the contaminants. In extreme cases, the contaminating atoms may form a solid solution in the magnesium and the silicon sublattices, as is the case for aluminum impurity in natural talc crystals. High-concentration doping can be used to modify the structural, optical, and electrical properties of talc. Note that scandium and aluminum can substitute both magnesium and silicon with low defect formation energies.

Throughout the discussion, we explored potential applications in photonics, spin qubits, and electronics, and identified promising \emph{candidate defects} for each of these areas. Interestingly, we found only one color center among the vacancy defects, the negatively charged silicon vacancy, that can give rise to a ZPL in the near-infrared region. The V$_{\text{Si}}$ defect may be realized on demand using electron and neutron irradiation, and the negative charge state is stable for Fermi energies close to the middle of the band gap.  Among the neutral magnesium substitutional defects, we identified possible optical transitions for the C$_{\text{Mg}}$ and Se$_{\text{Mg}}$ configurations. All of these defects, however, possess higher than 10~eV formation energy, thus they are not expected to be in high concentration. 
The only low formation energy color center we found is the positively charged  manganese substitutional defect emitting in the near-infrared region. Interestingly, we could not identify any color center for silicon substitutional defects in the talcum structure. In conclusion, due to the lack of probable photoactive defect, pure as-grown talc layers can be photo-inactive up to 7.27~eV photon energy, which makes it a suitable substrate for photonic applications, e.g.,  in van der Waals heterostructures. While this finding may seem discouraging at first glance, it actually sets the stage for engineering single-photon emitters. The absence of low-formation-energy defects and common impurity-related color centers suggests that one can expect minimal to no background emission from as-grown talc layers. Based on our results, manganese implantation appears to be the most promising candidate for realizing single-photon emitters in the telecom wavelength range. Such emitters may be introduced via ion implantation at an arbitrary low concentration. On the other hand, there may be many other transition metal and lanthanide impurities that are potentially interesting for single-photon emission. We did not explore this direction extensively, as the applied methodology, i.e.\ hybrid DFT with the HSE06 functional, may yield unreliable results due to the inaccurate description of the screening effects associated with d- and f-orbitals\cite{ivady_role_2013,ivady_theoretical_2014}. To uncover the full potential of transition metal and lanthanide impurities, further, more specialized studies are required.

In addition, we examined our results to reveal candidates for optically addressable spin qubits. In addition to an optical signal, spin qubits require high spin ($S > 1/2$) ground and optically excited state, and spin-selective nonradiative decays from the excited state through alternative spin states to the ground state. In the present study, which provides a broad overview of defects in talc, we were not able to verify the fulfillment of the latter condition. However, we were able to identify defects with high-spin ground states and optical activity, which make them \emph{promising candidates} for further investigation using more advanced methodologies \cite{benedek_accurate_2024}.
The most promising spin qubits are the Mn$_{\text{Mg}}^+$, the V$_{\text{Si}}^-$, and the Se$_{\text{Mg}}$. The latter two defects exhibit high formation energy, however; they may be realized by irradiation and implantation in low concentrations, facilitating single defect measurements. Owing to its low formation energy,  manganese-related spin qubit may be realized in high concentrations and lead to the realization of a 2D sensing sheet similar to  boron vacancy containing hBN\cite{tetienne_quantum_2021,healey_quantum_2022,kumar_magnetic_2022}. Here, we note that talcum possesses much fewer nuclear spins in its lattice than boron nitride. Thus we expect a longer coherence time for defect spins in talc.

We also comment on the charge stability of the potential spin qubits. In irradiated samples, the most abundant defects expected to form, due to their low formation energies, are hydrogen vacancies, -OH divacancies, and O(II.) vacancies. Given the likely high concentration of hydrogen vacancies, the Fermi level is expected to be pinned at approximately 3.01~eV above the valence band edge, which corresponds to the (0/–) charge transition level of V$_\text{H}$. At this Fermi level, the silicon vacancy is predicted to be negatively charged, see Fig.~\ref{fig:ctl}, which may implement a spin qubit. The other qubit candidate, the positively charged manganese substitutional defect, is expected to be stable in semi-insulating or p-type samples, see Fig.~\ref{fig:ctl}. The latter may be achieved through doping with sodium or by substituting aluminum with silicon.

Good acceptors and donor states in wide band gap semiconductors are very interesting for high-power electronic applications. Our study also shows that contaminated talcum layers can exhibit relatively shallow donor and acceptor states. Concerning the former, we showed that Al$_{\text{Mg}}$, P$_{\text{Si}}$, and potentially Sc$_{\text{Mg}}$ are promising candidates for realizing n-type talcum layers. In general, all of these impurities exhibit low formation. However, the Sc$_{\text{Mg}}$ impurity exhibits exceptionally low formation energy, suggesting a feasible path for manufacturing n-type talcum layers. Concerning acceptor states, Na$_{\text{Mg}}$, Al$_{\text{Si}}$, and Sc$_{\text{Si}}$ are the most promising dopants that exhibit low formation energy. Compared to the possible donors discussed before, the acceptor states move farther away from the valence band edge. While this may be detrimental to usual p-type conductivity, the low formation energy of the dopants opens new possibilities by enabling the formation of a defect band in highly doped talcum layers. 

In summary, we demonstrated through an extensive computational study that defective talcum layers can serve as the active layer of low-dimensional applications ranging from photonics through spin-based quantum sensing to electronics in low dimensions. Further experimental and theoretical studies are required to unveil the true potential of talcum.

\section{Methods}

To computationally predict the electronic structure and derived properties of talc, we employ density functional theory (DFT) calculations in periodic boundary conditions as implemented in the Vienna Ab initio Software Package (VASP)\cite{VASP,VASP2}. We use the projector-augmented wave (PAW) method\cite{PAW} to describe the potential of the nucleus and the weakly interacting core electrons of the atoms. Kohn-Sham wavefunctions of the valence electrons are expanded in a plane wave basis set of 440~eV for supercell models and 540~eV for the optimization of the unit cell. The stopping criteria for the electronic self-consistent field (SCF) loop and the structural relaxation are $\Delta E < 10^{-6}$~eV and $F_i < 10^{-3}$~eV/\AA , where $\Delta E$ is the change of the total energy in an iteration of the SCF loop, and $F_i$ is the length of the force vector acting on atom $i$. For unit cell optimization, we use an $8 \times 4 \times 4$ Gamma-centered k-point grid for sampling the Brillouin zone, while for the $3 \times 2 \times 2$ supercell of 504 atoms we use the Gamma-point only Brillouin zone sampling. To reduce the computational cost, we study single-layer models of 252 atoms for point defect calculations. Throughout the work, we use the HSE06 functional\cite{heyd_hybrid_2003,heyd_erratum:_2006} with the standard $\alpha = 0.25$ mixing parameter to approximately account for the exchange and correlation effects of the many-electron system. The untuned HSE06 functional is known to predict the bandgap of semiconductors with a low error margin.\cite{henderson_accurate_2011} To account for the weak van der Waals interaction between the layers, we employ the Grimme-D3 method\cite{grimme_consistent_2010} on top of the HSE06 exchange-correlation functional.
For the excited state calculations, we use the $\Delta$scf method\cite{gali_theory_2009} and promote one electron from the highest occupied Kohn-Sham defect orbital to the lowest unoccupied defect orbital in the spin channel of the lowest Kohn-Sham energy gap between the states. The excited state configurations have been relaxed as specified previously. For the charge transition level calculations, we correct the total energies of the charged supercells according to the jellium charge corrections method introduced in Ref.~[\cite{zhang_correcting_2023}].  For positively (negatively) charged single-layer 252-atom supercells, we obtain $E_{\text{corr}} = 0.044$~eV ($E_{\text{corr}} = 0.003$~eV). The difference in the magnitude of the corrections for +1 and -1 charge states is expectedly due to the difference in the dispersion of the conduction and valence band edges.
The valence band is flat, whereas the conduction band is not. Hyperfine tensor calculations were performed using the HSE06 functional and core polarization correction\cite{szasz_hyperfine_2013} in our 252-atom single-layer supercell model. In all relevant cases, we considered the most abundant spin-carrying isotope of each atom. For simplicity, we omitted the finite-size correction of the hyperfine interaction\cite{takacs_accurate_2024}, which introduces a relative uncertainty of 5–10\% in the computed tensor components, such as $A_{zz}$\cite{szasz_hyperfine_2013}.

The G$_0$W$_0$+BSE absorption spectrum calculation was carried out in VASP\cite{VASP}, using the HSE06-converged structure and wavefunction as a starting point. We estimate that a converged G$_0$W$_0$ bandgap is within the range of 9.1-9.2 eV. The BSE calculations were performed with the output data from the most converged GW calculation applying the Tamm-Dancoff approximation\cite{dancoffNonAdiabaticMesonTheory1950,tamm2011a}. For a detailed discussion of the calculations and results of the convergence test, see Supplementary Information.

\section*{Acknowledgments} 

This work was supported by the National Research, 
Development and Innovation Office of Hungary (NKFIH) within the Quantum Information National Laboratory of Hungary (Grant No. 2022-2.1.1-NL-2022-00004) and within the project FK 145395. 
V.I. and J.D. acknowledge support from the Knut and Alice Wallenberg Foundation through the WBSQD2 project (Grant No.\ 2018.0071). 
J.D. also acknowledges support from the Swedish Research Council (VR) Grant No. 2022-00276. The computations were enabled by resources provided by the National Academic Infrastructure for Supercomputing in Sweden (NAISS) at the Swedish National Infrastructure for Computing (SNIC) at Tetralith, partially funded by the Swedish Research Council through grant agreement no. 2022-06725.

\section*{Data availability}

The data that support the findings of this study are available from the authors upon request.

\section*{Code Availability}

The codes associated with this manuscript are available from the corresponding author upon request.

\section*{Author contributions}

G.D., V.I., and J.D.\ performed the DFT calculation, while O.B.L. performed the GW+BSE calculations.  The results were analyzed with contributions from all authors. V.I.\ conceived the idea of the project and supervised the work. The manuscript was written by V.I.\ and G.D.\ with inputs from the coauthors. 

\section*{Competing interests}

The authors declare no competing interests.


\begin{thebibliography}{50}%
\makeatletter
\providecommand \@ifxundefined [1]{%
 \@ifx{#1\undefined}
}%
\providecommand \@ifnum [1]{%
 \ifnum #1\expandafter \@firstoftwo
 \else \expandafter \@secondoftwo
 \fi
}%
\providecommand \@ifx [1]{%
 \ifx #1\expandafter \@firstoftwo
 \else \expandafter \@secondoftwo
 \fi
}%
\providecommand \natexlab [1]{#1}%
\providecommand \enquote  [1]{``#1''}%
\providecommand \bibnamefont  [1]{#1}%
\providecommand \bibfnamefont [1]{#1}%
\providecommand \citenamefont [1]{#1}%
\providecommand \href@noop [0]{\@secondoftwo}%
\providecommand \href [0]{\begingroup \@sanitize@url \@href}%
\providecommand \@href[1]{\@@startlink{#1}\@@href}%
\providecommand \@@href[1]{\endgroup#1\@@endlink}%
\providecommand \@sanitize@url [0]{\catcode `\\12\catcode `\$12\catcode
  `\&12\catcode `\#12\catcode `\^12\catcode `\_12\catcode `\%12\relax}%
\providecommand \@@startlink[1]{}%
\providecommand \@@endlink[0]{}%
\providecommand \url  [0]{\begingroup\@sanitize@url \@url }%
\providecommand \@url [1]{\endgroup\@href {#1}{\urlprefix }}%
\providecommand \urlprefix  [0]{URL }%
\providecommand \Eprint [0]{\href }%
\providecommand \doibase [0]{http://dx.doi.org/}%
\providecommand \selectlanguage [0]{\@gobble}%
\providecommand \bibinfo  [0]{\@secondoftwo}%
\providecommand \bibfield  [0]{\@secondoftwo}%
\providecommand \translation [1]{[#1]}%
\providecommand \BibitemOpen [0]{}%
\providecommand \bibitemStop [0]{}%
\providecommand \bibitemNoStop [0]{.\EOS\space}%
\providecommand \EOS [0]{\spacefactor3000\relax}%
\providecommand \BibitemShut  [1]{\csname bibitem#1\endcsname}%
\let\auto@bib@innerbib\@empty
\bibitem [{\citenamefont {Geim}\ and\ \citenamefont
  {Grigorieva}(2013)}]{geim_van_2013}%
  \BibitemOpen
  \bibfield  {author} {\bibinfo {author} {\bibfnamefont {A.~K.}\ \bibnamefont
  {Geim}}\ and\ \bibinfo {author} {\bibfnamefont {I.~V.}\ \bibnamefont
  {Grigorieva}},\ }\href {\doibase 10.1038/nature12385} {\bibfield  {journal}
  {\bibinfo  {journal} {Nature}\ }\textbf {\bibinfo {volume} {499}},\ \bibinfo
  {pages} {419} (\bibinfo {year} {2013})},\ \bibinfo {note} {number: 7459
  Publisher: Nature Publishing Group}\BibitemShut {NoStop}%
\bibitem [{\citenamefont {Liu}\ \emph {et~al.}(2016)\citenamefont {Liu},
  \citenamefont {Weiss}, \citenamefont {Duan}, \citenamefont {Cheng},
  \citenamefont {Huang},\ and\ \citenamefont {Duan}}]{liu_van_2016}%
  \BibitemOpen
  \bibfield  {author} {\bibinfo {author} {\bibfnamefont {Y.}~\bibnamefont
  {Liu}}, \bibinfo {author} {\bibfnamefont {N.~O.}\ \bibnamefont {Weiss}},
  \bibinfo {author} {\bibfnamefont {X.}~\bibnamefont {Duan}}, \bibinfo {author}
  {\bibfnamefont {H.-C.}\ \bibnamefont {Cheng}}, \bibinfo {author}
  {\bibfnamefont {Y.}~\bibnamefont {Huang}}, \ and\ \bibinfo {author}
  {\bibfnamefont {X.}~\bibnamefont {Duan}},\ }\href {\doibase
  10.1038/natrevmats.2016.42} {\bibfield  {journal} {\bibinfo  {journal}
  {Nature Reviews Materials}\ }\textbf {\bibinfo {volume} {1}},\ \bibinfo
  {pages} {1} (\bibinfo {year} {2016})},\ \bibinfo {note} {publisher: Nature
  Publishing Group}\BibitemShut {NoStop}%
\bibitem [{\citenamefont {Liang}\ \emph {et~al.}(2020)\citenamefont {Liang},
  \citenamefont {Cheng}, \citenamefont {Cui},\ and\ \citenamefont
  {Miao}}]{liang_van_2020}%
  \BibitemOpen
  \bibfield  {author} {\bibinfo {author} {\bibfnamefont {S.-J.}\ \bibnamefont
  {Liang}}, \bibinfo {author} {\bibfnamefont {B.}~\bibnamefont {Cheng}},
  \bibinfo {author} {\bibfnamefont {X.}~\bibnamefont {Cui}}, \ and\ \bibinfo
  {author} {\bibfnamefont {F.}~\bibnamefont {Miao}},\ }\href {\doibase
  10.1002/adma.201903800} {\bibfield  {journal} {\bibinfo  {journal} {Advanced
  Materials}\ }\textbf {\bibinfo {volume} {32}},\ \bibinfo {pages} {1903800}
  (\bibinfo {year} {2020})} \BibitemShut
  {NoStop}%
\bibitem [{\citenamefont {Healey}\ \emph {et~al.}(2022)\citenamefont {Healey},
  \citenamefont {Scholten}, \citenamefont {Yang}, \citenamefont {Scott},
  \citenamefont {Abrahams}, \citenamefont {Robertson}, \citenamefont {Hou},
  \citenamefont {Guo}, \citenamefont {Rahman}, \citenamefont {Lu},
  \citenamefont {Kianinia}, \citenamefont {Aharonovich},\ and\ \citenamefont
  {Tetienne}}]{healey_quantum_2022}%
  \BibitemOpen
  \bibfield  {author} {\bibinfo {author} {\bibfnamefont {A.~J.}\ \bibnamefont
  {Healey}}, \bibinfo {author} {\bibfnamefont {S.~C.}\ \bibnamefont
  {Scholten}}, \bibinfo {author} {\bibfnamefont {T.}~\bibnamefont {Yang}},
  \bibinfo {author} {\bibfnamefont {J.~A.}\ \bibnamefont {Scott}}, \bibinfo
  {author} {\bibfnamefont {G.~J.}\ \bibnamefont {Abrahams}}, \bibinfo {author}
  {\bibfnamefont {I.~O.}\ \bibnamefont {Robertson}}, \bibinfo {author}
  {\bibfnamefont {X.~F.}\ \bibnamefont {Hou}}, \bibinfo {author} {\bibfnamefont
  {Y.~F.}\ \bibnamefont {Guo}}, \bibinfo {author} {\bibfnamefont
  {S.}~\bibnamefont {Rahman}}, \bibinfo {author} {\bibfnamefont
  {Y.}~\bibnamefont {Lu}}, \bibinfo {author} {\bibfnamefont {M.}~\bibnamefont
  {Kianinia}}, \bibinfo {author} {\bibfnamefont {I.}~\bibnamefont
  {Aharonovich}}, \ and\ \bibinfo {author} {\bibfnamefont {J.-P.}\ \bibnamefont
  {Tetienne}},\ }\href {\doibase 10.1038/s41567-022-01815-5} {\bibfield
  {journal} {\bibinfo  {journal} {Nature Physics}\ ,\ \bibinfo {pages} {1}}
  (\bibinfo {year} {2022})},\ \bibinfo {note} {publisher: Nature Publishing
  Group}\BibitemShut {NoStop}%
\bibitem [{\citenamefont {Frisenda}\ \emph {et~al.}(2020)\citenamefont
  {Frisenda}, \citenamefont {Niu}, \citenamefont {Gant}, \citenamefont
  {Munoz},\ and\ \citenamefont {Castellanos-Gomez}}]{frisenda_naturally_2020}%
  \BibitemOpen
  \bibfield  {author} {\bibinfo {author} {\bibfnamefont {R.}~\bibnamefont
  {Frisenda}}, \bibinfo {author} {\bibfnamefont {Y.}~\bibnamefont {Niu}},
  \bibinfo {author} {\bibfnamefont {P.}~\bibnamefont {Gant}}, \bibinfo {author}
  {\bibfnamefont {M.}~\bibnamefont {Munoz}}, \ and\ \bibinfo {author}
  {\bibfnamefont {A.}~\bibnamefont {Castellanos-Gomez}},\ }\href {\doibase
  10.1038/s41699-020-00172-2} {\bibfield  {journal} {\bibinfo  {journal} {npj
  2D Materials and Applications}\ }\textbf {\bibinfo {volume} {4}},\ \bibinfo
  {pages} {1} (\bibinfo {year} {2020})},\ \bibinfo {note} {publisher: Nature
  Publishing Group}\BibitemShut {NoStop}%
\bibitem [{\citenamefont {Claverie}\ \emph {et~al.}(2018)\citenamefont
  {Claverie}, \citenamefont {Dumas}, \citenamefont {Careme}, \citenamefont
  {Poirier}, \citenamefont {Le~Roux}, \citenamefont {Micoud}, \citenamefont
  {Martin},\ and\ \citenamefont {Aymonier}}]{claverie_synthetic_2018}%
  \BibitemOpen
  \bibfield  {author} {\bibinfo {author} {\bibfnamefont {M.}~\bibnamefont
  {Claverie}}, \bibinfo {author} {\bibfnamefont {A.}~\bibnamefont {Dumas}},
  \bibinfo {author} {\bibfnamefont {C.}~\bibnamefont {Careme}}, \bibinfo
  {author} {\bibfnamefont {M.}~\bibnamefont {Poirier}}, \bibinfo {author}
  {\bibfnamefont {C.}~\bibnamefont {Le~Roux}}, \bibinfo {author} {\bibfnamefont
  {P.}~\bibnamefont {Micoud}}, \bibinfo {author} {\bibfnamefont
  {F.}~\bibnamefont {Martin}}, \ and\ \bibinfo {author} {\bibfnamefont
  {C.}~\bibnamefont {Aymonier}},\ }\href {\doibase 10.1002/chem.201702763}
  {\bibfield  {journal} {\bibinfo  {journal} {Chemistry‚ A European
  Journal}\ }\textbf {\bibinfo {volume} {24}},\ \bibinfo {pages} {519}
  (\bibinfo {year} {2018})} \BibitemShut
  {NoStop}%
\bibitem [{\citenamefont {Vasiƒá}\ \emph {et~al.}(2021)\citenamefont {Vasiƒá},
  \citenamefont {Czibula}, \citenamefont {Kratzer}, \citenamefont {R~A~Neves},
  \citenamefont {Matkoviƒá},\ and\ \citenamefont
  {Teichert}}]{vasic_two-dimensional_2021}%
  \BibitemOpen
  \bibfield  {author} {\bibinfo {author} {\bibfnamefont {B.}~\bibnamefont
  {Vasiƒá}}, \bibinfo {author} {\bibfnamefont {C.}~\bibnamefont {Czibula}},
  \bibinfo {author} {\bibfnamefont {M.}~\bibnamefont {Kratzer}}, \bibinfo
  {author} {\bibfnamefont {B.}~\bibnamefont {R~A~Neves}}, \bibinfo {author}
  {\bibfnamefont {A.}~\bibnamefont {Matkoviƒá}}, \ and\ \bibinfo {author}
  {\bibfnamefont {C.}~\bibnamefont {Teichert}},\ }\href {\doibase
  10.1088/1361-6528/abeffe} {\bibfield  {journal} {\bibinfo  {journal}
  {Nanotechnology}\ }\textbf {\bibinfo {volume} {32}},\ \bibinfo {pages}
  {265701} (\bibinfo {year} {2021})}\BibitemShut {NoStop}%
\bibitem [{\citenamefont {Mania}\ \emph {et~al.}(2017)\citenamefont {Mania},
  \citenamefont {Alencar}, \citenamefont {Cadore}, \citenamefont {Carvalho},
  \citenamefont {Watanabe}, \citenamefont {Taniguchi}, \citenamefont {Neves},
  \citenamefont {Chacham},\ and\ \citenamefont
  {Campos}}]{mania_spontaneous_2017}%
  \BibitemOpen
  \bibfield  {author} {\bibinfo {author} {\bibfnamefont {E.}~\bibnamefont
  {Mania}}, \bibinfo {author} {\bibfnamefont {A.~B.}\ \bibnamefont {Alencar}},
  \bibinfo {author} {\bibfnamefont {A.~R.}\ \bibnamefont {Cadore}}, \bibinfo
  {author} {\bibfnamefont {B.~R.}\ \bibnamefont {Carvalho}}, \bibinfo {author}
  {\bibfnamefont {K.}~\bibnamefont {Watanabe}}, \bibinfo {author}
  {\bibfnamefont {T.}~\bibnamefont {Taniguchi}}, \bibinfo {author}
  {\bibfnamefont {B.~R.~A.}\ \bibnamefont {Neves}}, \bibinfo {author}
  {\bibfnamefont {H.}~\bibnamefont {Chacham}}, \ and\ \bibinfo {author}
  {\bibfnamefont {L.~C.}\ \bibnamefont {Campos}},\ }\href {\doibase
  10.1088/2053-1583/aa76f4} {\bibfield  {journal} {\bibinfo  {journal} {2D
  Materials}\ }\textbf {\bibinfo {volume} {4}},\ \bibinfo {pages} {031008}
  (\bibinfo {year} {2017})},\ \bibinfo {note} {publisher: IOP
  Publishing}\BibitemShut {NoStop}%
\bibitem [{\citenamefont {Barcelos}\ \emph {et~al.}(2018)\citenamefont
  {Barcelos}, \citenamefont {Cadore}, \citenamefont {Alencar}, \citenamefont
  {Maia}, \citenamefont {Mania}, \citenamefont {Oliveira}, \citenamefont
  {Bufon}, \citenamefont {Malachias}, \citenamefont {Freitas}, \citenamefont
  {Moreira},\ and\ \citenamefont {Chacham}}]{barcelos_infrared_2018}%
  \BibitemOpen
  \bibfield  {author} {\bibinfo {author} {\bibfnamefont {I.~D.}\ \bibnamefont
  {Barcelos}}, \bibinfo {author} {\bibfnamefont {A.~R.}\ \bibnamefont
  {Cadore}}, \bibinfo {author} {\bibfnamefont {A.~B.}\ \bibnamefont {Alencar}},
  \bibinfo {author} {\bibfnamefont {F.~C.~B.}\ \bibnamefont {Maia}}, \bibinfo
  {author} {\bibfnamefont {E.}~\bibnamefont {Mania}}, \bibinfo {author}
  {\bibfnamefont {R.~F.}\ \bibnamefont {Oliveira}}, \bibinfo {author}
  {\bibfnamefont {C.~C.~B.}\ \bibnamefont {Bufon}}, \bibinfo {author}
  {\bibfnamefont {A.}~\bibnamefont {Malachias}}, \bibinfo {author}
  {\bibfnamefont {R.~O.}\ \bibnamefont {Freitas}}, \bibinfo {author}
  {\bibfnamefont {R.~L.}\ \bibnamefont {Moreira}}, \ and\ \bibinfo {author}
  {\bibfnamefont {H.}~\bibnamefont {Chacham}},\ }\href {\doibase
  10.1021/acsphotonics.7b01017} {\bibfield  {journal} {\bibinfo  {journal} {ACS
  Photonics}\ }\textbf {\bibinfo {volume} {5}},\ \bibinfo {pages} {1912}
  (\bibinfo {year} {2018})},\ \bibinfo {note} {publisher: American Chemical
  Society}\BibitemShut {NoStop}%
\bibitem [{\citenamefont {Gadelha}\ \emph {et~al.}(2021)\citenamefont
  {Gadelha}, \citenamefont {Vasconcelos}, \citenamefont {Can√ßado},\ and\
  \citenamefont {Jorio}}]{gadelha_nano-optical_2021}%
  \BibitemOpen
  \bibfield  {author} {\bibinfo {author} {\bibfnamefont {A.~C.}\ \bibnamefont
  {Gadelha}}, \bibinfo {author} {\bibfnamefont {T.~L.}\ \bibnamefont
  {Vasconcelos}}, \bibinfo {author} {\bibfnamefont {L.~G.}\ \bibnamefont
  {Cancado}}, \ and\ \bibinfo {author} {\bibfnamefont {A.}~\bibnamefont
  {Jorio}},\ }\href {\doibase 10.1021/acs.jpclett.1c01804} {\bibfield
  {journal} {\bibinfo  {journal} {The Journal of Physical Chemistry Letters}\
  }\textbf {\bibinfo {volume} {12}},\ \bibinfo {pages} {7625} (\bibinfo {year}
  {2021})},\ \bibinfo {note} {publisher: American Chemical Society}\BibitemShut
  {NoStop}%
\bibitem [{\citenamefont {Prando}\ \emph {et~al.}(2021)\citenamefont {Prando},
  \citenamefont {Severijnen}, \citenamefont {Barcelos}, \citenamefont
  {Zeitler}, \citenamefont {Christianen}, \citenamefont {Withers},\ and\
  \citenamefont {Galv√£o~Gobato}}]{prando_revealing_2021}%
  \BibitemOpen
  \bibfield  {author} {\bibinfo {author} {\bibfnamefont {G.~A.}\ \bibnamefont
  {Prando}}, \bibinfo {author} {\bibfnamefont {M.~E.}\ \bibnamefont
  {Severijnen}}, \bibinfo {author} {\bibfnamefont {I.~D.}\ \bibnamefont
  {Barcelos}}, \bibinfo {author} {\bibfnamefont {U.}~\bibnamefont {Zeitler}},
  \bibinfo {author} {\bibfnamefont {P.~C.~M.}\ \bibnamefont {Christianen}},
  \bibinfo {author} {\bibfnamefont {F.}~\bibnamefont {Withers}}, \ and\
  \bibinfo {author} {\bibfnamefont {Y.}~\bibnamefont {Galvao~Gobato}},\ }\href
  {\doibase 10.1103/PhysRevApplied.16.064055} {\bibfield  {journal} {\bibinfo
  {journal} {Physical Review Applied}\ }\textbf {\bibinfo {volume} {16}},\
  \bibinfo {pages} {064055} (\bibinfo {year} {2021})},\ \bibinfo {note}
  {publisher: American Physical Society}\BibitemShut {NoStop}%
\bibitem [{\citenamefont {Nutting}\ \emph {et~al.}(2021)\citenamefont
  {Nutting}, \citenamefont {A.~Prando}, \citenamefont {Severijnen},
  \citenamefont {D.~Barcelos}, \citenamefont {Guo}, \citenamefont
  {M.~Christianen}, \citenamefont {Zeitler}, \citenamefont {Gobato},\ and\
  \citenamefont {Withers}}]{nutting_electrical_2021}%
  \BibitemOpen
  \bibfield  {author} {\bibinfo {author} {\bibfnamefont {D.}~\bibnamefont
  {Nutting}}, \bibinfo {author} {\bibfnamefont {G.}~\bibnamefont {A.~Prando}},
  \bibinfo {author} {\bibfnamefont {M.}~\bibnamefont {Severijnen}}, \bibinfo
  {author} {\bibfnamefont {I.}~\bibnamefont {D.~Barcelos}}, \bibinfo {author}
  {\bibfnamefont {S.}~\bibnamefont {Guo}}, \bibinfo {author} {\bibfnamefont
  {P.~C.}\ \bibnamefont {M.~Christianen}}, \bibinfo {author} {\bibfnamefont
  {U.}~\bibnamefont {Zeitler}}, \bibinfo {author} {\bibfnamefont {Y.~G.}\
  \bibnamefont {Gobato}}, \ and\ \bibinfo {author} {\bibfnamefont
  {F.}~\bibnamefont {Withers}},\ }\href {\doibase 10.1039/D1NR04723J}
  {\bibfield  {journal} {\bibinfo  {journal} {Nanoscale}\ }\textbf {\bibinfo
  {volume} {13}},\ \bibinfo {pages} {15853} (\bibinfo {year} {2021})},\
  \bibinfo {note} {publisher: Royal Society of Chemistry}\BibitemShut {NoStop}%
\bibitem [{\citenamefont {Costa}\ \emph {et~al.}(2023)\citenamefont {Costa},
  \citenamefont {Brito}, \citenamefont {Barcelos},\ and\ \citenamefont
  {Zagonel}}]{costa_impacts_2023}%
  \BibitemOpen
  \bibfield  {author} {\bibinfo {author} {\bibfnamefont {F.~J.~R.}\
  \bibnamefont {Costa}}, \bibinfo {author} {\bibfnamefont {T.~G.-L.}\
  \bibnamefont {Brito}}, \bibinfo {author} {\bibfnamefont {I.~D.}\ \bibnamefont
  {Barcelos}}, \ and\ \bibinfo {author} {\bibfnamefont {L.~F.}\ \bibnamefont
  {Zagonel}},\ }\href {\doibase 10.1088/1361-6528/acda3b} {\bibfield  {journal}
  {\bibinfo  {journal} {Nanotechnology}\ }\textbf {\bibinfo {volume} {34}},\
  \bibinfo {pages} {385703} (\bibinfo {year} {2023})},\ \bibinfo {note}
  {publisher: IOP Publishing}\BibitemShut {NoStop}%
\bibitem [{\citenamefont {Alencar}\ \emph {et~al.}(2015)\citenamefont
  {Alencar}, \citenamefont {Barboza}, \citenamefont {Archanjo}, \citenamefont
  {Chacham},\ and\ \citenamefont {Neves}}]{alencar_experimental_2015}%
  \BibitemOpen
  \bibfield  {author} {\bibinfo {author} {\bibfnamefont {A.~B.}\ \bibnamefont
  {Alencar}}, \bibinfo {author} {\bibfnamefont {A.~P.~M.}\ \bibnamefont
  {Barboza}}, \bibinfo {author} {\bibfnamefont {B.~S.}\ \bibnamefont
  {Archanjo}}, \bibinfo {author} {\bibfnamefont {H.}~\bibnamefont {Chacham}}, \
  and\ \bibinfo {author} {\bibfnamefont {B.~R.~A.}\ \bibnamefont {Neves}},\
  }\href {\doibase 10.1088/2053-1583/2/1/015004} {\bibfield  {journal}
  {\bibinfo  {journal} {2D Materials}\ }\textbf {\bibinfo {volume} {2}},\
  \bibinfo {pages} {015004} (\bibinfo {year} {2015})}\BibitemShut {NoStop}%
\bibitem [{\citenamefont {Ma}\ \emph {et~al.}(2022)\citenamefont {Ma},
  \citenamefont {Du}, \citenamefont {Lan}, \citenamefont {Chen},\ and\
  \citenamefont {Lan}}]{ma_effect_2022}%
  \BibitemOpen
  \bibfield  {author} {\bibinfo {author} {\bibfnamefont {X.}~\bibnamefont
  {Ma}}, \bibinfo {author} {\bibfnamefont {H.}~\bibnamefont {Du}}, \bibinfo
  {author} {\bibfnamefont {P.}~\bibnamefont {Lan}}, \bibinfo {author}
  {\bibfnamefont {J.}~\bibnamefont {Chen}}, \ and\ \bibinfo {author}
  {\bibfnamefont {L.}~\bibnamefont {Lan}},\ }\href {\doibase
  10.3390/min12010069} {\bibfield  {journal} {\bibinfo  {journal} {Minerals}\
  }\textbf {\bibinfo {volume} {12}},\ \bibinfo {pages} {69} (\bibinfo {year}
  {2022})},\ \bibinfo {note} {number: 1 Publisher: Multidisciplinary Digital
  Publishing Institute}\BibitemShut {NoStop}%
\bibitem [{\citenamefont {Matkoviƒá}\ \emph {et~al.}(2021)\citenamefont
  {Matkoviƒá}, \citenamefont {Ludescher}, \citenamefont {Peil}, \citenamefont
  {Sharma}, \citenamefont {Gradwohl}, \citenamefont {Kratzer}, \citenamefont
  {Zimmermann}, \citenamefont {Genser}, \citenamefont {Knez}, \citenamefont
  {Fisslthaler}, \citenamefont {Gammer}, \citenamefont {Lugstein},
  \citenamefont {Bakker}, \citenamefont {Romaner}, \citenamefont {Zahn},
  \citenamefont {Hofer}, \citenamefont {Salvan}, \citenamefont {Raith},\ and\
  \citenamefont {Teichert}}]{matkovic_iron-rich_2021}%
  \BibitemOpen
  \bibfield  {author} {\bibinfo {author} {\bibfnamefont {A.}~\bibnamefont
  {Matkoviƒá}}, \bibinfo {author} {\bibfnamefont {L.}~\bibnamefont
  {Ludescher}}, \bibinfo {author} {\bibfnamefont {O.~E.}\ \bibnamefont {Peil}},
  \bibinfo {author} {\bibfnamefont {A.}~\bibnamefont {Sharma}}, \bibinfo
  {author} {\bibfnamefont {K.-P.}\ \bibnamefont {Gradwohl}}, \bibinfo {author}
  {\bibfnamefont {M.}~\bibnamefont {Kratzer}}, \bibinfo {author} {\bibfnamefont
  {M.}~\bibnamefont {Zimmermann}}, \bibinfo {author} {\bibfnamefont
  {J.}~\bibnamefont {Genser}}, \bibinfo {author} {\bibfnamefont
  {D.}~\bibnamefont {Knez}}, \bibinfo {author} {\bibfnamefont {E.}~\bibnamefont
  {Fisslthaler}}, \bibinfo {author} {\bibfnamefont {C.}~\bibnamefont {Gammer}},
  \bibinfo {author} {\bibfnamefont {A.}~\bibnamefont {Lugstein}}, \bibinfo
  {author} {\bibfnamefont {R.~J.}\ \bibnamefont {Bakker}}, \bibinfo {author}
  {\bibfnamefont {L.}~\bibnamefont {Romaner}}, \bibinfo {author} {\bibfnamefont
  {D.~R.~T.}\ \bibnamefont {Zahn}}, \bibinfo {author} {\bibfnamefont
  {F.}~\bibnamefont {Hofer}}, \bibinfo {author} {\bibfnamefont
  {G.}~\bibnamefont {Salvan}}, \bibinfo {author} {\bibfnamefont {J.~G.}\
  \bibnamefont {Raith}}, \ and\ \bibinfo {author} {\bibfnamefont
  {C.}~\bibnamefont {Teichert}},\ }\href {\doibase 10.1038/s41699-021-00276-3}
  {\bibfield  {journal} {\bibinfo  {journal} {npj 2D Materials and
  Applications}\ }\textbf {\bibinfo {volume} {5}},\ \bibinfo {pages} {1}
  (\bibinfo {year} {2021})},\ \bibinfo {note} {publisher: Nature Publishing
  Group}\BibitemShut {NoStop}%
\bibitem [{\citenamefont {Aharonovich}\ \emph {et~al.}(2016)\citenamefont
  {Aharonovich}, \citenamefont {Englund},\ and\ \citenamefont
  {Toth}}]{aharonovich_solid-state_2016}%
  \BibitemOpen
  \bibfield  {author} {\bibinfo {author} {\bibfnamefont {I.}~\bibnamefont
  {Aharonovich}}, \bibinfo {author} {\bibfnamefont {D.}~\bibnamefont
  {Englund}}, \ and\ \bibinfo {author} {\bibfnamefont {M.}~\bibnamefont
  {Toth}},\ }\href {\doibase 10.1038/nphoton.2016.186} {\bibfield  {journal}
  {\bibinfo  {journal} {Nature Photonics}\ }\textbf {\bibinfo {volume} {10}},\
  \bibinfo {pages} {631} (\bibinfo {year} {2016})},\ \bibinfo {note} {number:
  10 Publisher: Nature Publishing Group}\BibitemShut {NoStop}%
\bibitem [{\citenamefont {Awschalom}\ \emph {et~al.}(2018)\citenamefont
  {Awschalom}, \citenamefont {Hanson}, \citenamefont {Wrachtrup},\ and\
  \citenamefont {Zhou}}]{awschalom_quantum_2018}%
  \BibitemOpen
  \bibfield  {author} {\bibinfo {author} {\bibfnamefont {D.~D.}\ \bibnamefont
  {Awschalom}}, \bibinfo {author} {\bibfnamefont {R.}~\bibnamefont {Hanson}},
  \bibinfo {author} {\bibfnamefont {J.}~\bibnamefont {Wrachtrup}}, \ and\
  \bibinfo {author} {\bibfnamefont {B.~B.}\ \bibnamefont {Zhou}},\ }\href
  {\doibase 10.1038/s41566-018-0232-2} {\bibfield  {journal} {\bibinfo
  {journal} {Nature Photonics}\ }\textbf {\bibinfo {volume} {12}},\ \bibinfo
  {pages} {516} (\bibinfo {year} {2018})},\ \bibinfo {note} {number: 9
  Publisher: Nature Publishing Group}\BibitemShut {NoStop}%
\bibitem [{\citenamefont {Schirhagl}\ \emph {et~al.}(2014)\citenamefont
  {Schirhagl}, \citenamefont {Chang}, \citenamefont {Loretz},\ and\
  \citenamefont {Degen}}]{schirhagl_nitrogen-vacancy_2014}%
  \BibitemOpen
  \bibfield  {author} {\bibinfo {author} {\bibfnamefont {R.}~\bibnamefont
  {Schirhagl}}, \bibinfo {author} {\bibfnamefont {K.}~\bibnamefont {Chang}},
  \bibinfo {author} {\bibfnamefont {M.}~\bibnamefont {Loretz}}, \ and\ \bibinfo
  {author} {\bibfnamefont {C.~L.}\ \bibnamefont {Degen}},\ }\href {\doibase
  10.1146/annurev-physchem-040513-103659} {\bibfield  {journal} {\bibinfo
  {journal} {Annual Review of Physical Chemistry}\ }\textbf {\bibinfo {volume}
  {65}},\ \bibinfo {pages} {83} (\bibinfo {year} {2014})} \BibitemShut {NoStop}%
\bibitem [{\citenamefont {Gottscholl}\ \emph {et~al.}(2020)\citenamefont
  {Gottscholl}, \citenamefont {Kianinia}, \citenamefont {Soltamov},
  \citenamefont {Orlinskii}, \citenamefont {Mamin}, \citenamefont {Bradac},
  \citenamefont {Kasper}, \citenamefont {Krambrock}, \citenamefont {Sperlich},
  \citenamefont {Toth}, \citenamefont {Aharonovich},\ and\ \citenamefont
  {Dyakonov}}]{gottscholl_initialization_2020}%
  \BibitemOpen
  \bibfield  {author} {\bibinfo {author} {\bibfnamefont {A.}~\bibnamefont
  {Gottscholl}}, \bibinfo {author} {\bibfnamefont {M.}~\bibnamefont
  {Kianinia}}, \bibinfo {author} {\bibfnamefont {V.}~\bibnamefont {Soltamov}},
  \bibinfo {author} {\bibfnamefont {S.}~\bibnamefont {Orlinskii}}, \bibinfo
  {author} {\bibfnamefont {G.}~\bibnamefont {Mamin}}, \bibinfo {author}
  {\bibfnamefont {C.}~\bibnamefont {Bradac}}, \bibinfo {author} {\bibfnamefont
  {C.}~\bibnamefont {Kasper}}, \bibinfo {author} {\bibfnamefont
  {K.}~\bibnamefont {Krambrock}}, \bibinfo {author} {\bibfnamefont
  {A.}~\bibnamefont {Sperlich}}, \bibinfo {author} {\bibfnamefont
  {M.}~\bibnamefont {Toth}}, \bibinfo {author} {\bibfnamefont {I.}~\bibnamefont
  {Aharonovich}}, \ and\ \bibinfo {author} {\bibfnamefont {V.}~\bibnamefont
  {Dyakonov}},\ }\href {\doibase 10.1038/s41563-020-0619-6} {\bibfield
  {journal} {\bibinfo  {journal} {Nature Materials}\ }\textbf {\bibinfo
  {volume} {19}},\ \bibinfo {pages} {540} (\bibinfo {year} {2020})},\ \bibinfo
  {note} {number: 5 Publisher: Nature Publishing Group}\BibitemShut {NoStop}%
\bibitem [{\citenamefont {Iv√°dy}\ \emph {et~al.}(2020)\citenamefont {Iv√°dy},
  \citenamefont {Barcza}, \citenamefont {Thiering}, \citenamefont {Li},
  \citenamefont {Hamdi}, \citenamefont {Chou}, \citenamefont {Legeza},\ and\
  \citenamefont {Gali}}]{ivady_ab_2020}%
  \BibitemOpen
  \bibfield  {author} {\bibinfo {author} {\bibfnamefont {V.}~\bibnamefont
  {Ivády}}, \bibinfo {author} {\bibfnamefont {G.}~\bibnamefont {Barcza}},
  \bibinfo {author} {\bibfnamefont {G.}~\bibnamefont {Thiering}}, \bibinfo
  {author} {\bibfnamefont {S.}~\bibnamefont {Li}}, \bibinfo {author}
  {\bibfnamefont {H.}~\bibnamefont {Hamdi}}, \bibinfo {author} {\bibfnamefont
  {J.-P.}\ \bibnamefont {Chou}}, \bibinfo {author} {\bibfnamefont
  {Ö.}~\bibnamefont {Legeza}}, \ and\ \bibinfo {author} {\bibfnamefont
  {A.}~\bibnamefont {Gali}},\ }\href {\doibase 10.1038/s41524-020-0305-x}
  {\bibfield  {journal} {\bibinfo  {journal} {npj Computational Materials}\
  }\textbf {\bibinfo {volume} {6}},\ \bibinfo {pages} {1} (\bibinfo {year}
  {2020})},\ \bibinfo {note} {number: 1 Publisher: Nature Publishing
  Group}\BibitemShut {NoStop}%
\bibitem [{\citenamefont {Stern}\ \emph {et~al.}(2024)\citenamefont {Stern},
  \citenamefont {M.~Gilardoni}, \citenamefont {Gu}, \citenamefont
  {Eizagirre~Barker}, \citenamefont {Powell}, \citenamefont {Deng},
  \citenamefont {Fraser}, \citenamefont {Follet}, \citenamefont {Li},
  \citenamefont {Ramsay}, \citenamefont {Tan}, \citenamefont {Aharonovich},\
  and\ \citenamefont {Atatüre}}]{stern_quantum_2024}%
  \BibitemOpen
  \bibfield  {author} {\bibinfo {author} {\bibfnamefont {H.~L.}\ \bibnamefont
  {Stern}}, \bibinfo {author} {\bibfnamefont {C.}~\bibnamefont {M.~Gilardoni}},
  \bibinfo {author} {\bibfnamefont {Q.}~\bibnamefont {Gu}}, \bibinfo {author}
  {\bibfnamefont {S.}~\bibnamefont {Eizagirre~Barker}}, \bibinfo {author}
  {\bibfnamefont {O.~F.~J.}\ \bibnamefont {Powell}}, \bibinfo {author}
  {\bibfnamefont {X.}~\bibnamefont {Deng}}, \bibinfo {author} {\bibfnamefont
  {S.~A.}\ \bibnamefont {Fraser}}, \bibinfo {author} {\bibfnamefont
  {L.}~\bibnamefont {Follet}}, \bibinfo {author} {\bibfnamefont
  {C.}~\bibnamefont {Li}}, \bibinfo {author} {\bibfnamefont {A.~J.}\
  \bibnamefont {Ramsay}}, \bibinfo {author} {\bibfnamefont {H.~H.}\
  \bibnamefont {Tan}}, \bibinfo {author} {\bibfnamefont {I.}~\bibnamefont
  {Aharonovich}}, \ and\ \bibinfo {author} {\bibfnamefont {M.}~\bibnamefont
  {Atatüre}},\ }\href {\doibase 10.1038/s41563-024-01887-z} {\bibfield
  {journal} {\bibinfo  {journal} {Nature Materials}\ ,\ \bibinfo {pages} {1}}
  (\bibinfo {year} {2024})},\ \bibinfo {note} {publisher: Nature Publishing
  Group}\BibitemShut {NoStop}%
\bibitem [{\citenamefont {Kumar}\ \emph {et~al.}(2022)\citenamefont {Kumar},
  \citenamefont {Fabre}, \citenamefont {Durand}, \citenamefont {Clua-Provost},
  \citenamefont {Li}, \citenamefont {Edgar}, \citenamefont {Rougemaille},
  \citenamefont {Coraux}, \citenamefont {Marie}, \citenamefont {Renucci},
  \citenamefont {Robert}, \citenamefont {Robert-Philip}, \citenamefont {Gil},
  \citenamefont {Cassabois}, \citenamefont {Finco},\ and\ \citenamefont
  {Jacques}}]{kumar_magnetic_2022}%
  \BibitemOpen
  \bibfield  {author} {\bibinfo {author} {\bibfnamefont {P.}~\bibnamefont
  {Kumar}}, \bibinfo {author} {\bibfnamefont {F.}~\bibnamefont {Fabre}},
  \bibinfo {author} {\bibfnamefont {A.}~\bibnamefont {Durand}}, \bibinfo
  {author} {\bibfnamefont {T.}~\bibnamefont {Clua-Provost}}, \bibinfo {author}
  {\bibfnamefont {J.}~\bibnamefont {Li}}, \bibinfo {author} {\bibfnamefont
  {J.}~\bibnamefont {Edgar}}, \bibinfo {author} {\bibfnamefont
  {N.}~\bibnamefont {Rougemaille}}, \bibinfo {author} {\bibfnamefont
  {J.}~\bibnamefont {Coraux}}, \bibinfo {author} {\bibfnamefont
  {X.}~\bibnamefont {Marie}}, \bibinfo {author} {\bibfnamefont
  {P.}~\bibnamefont {Renucci}}, \bibinfo {author} {\bibfnamefont
  {C.}~\bibnamefont {Robert}}, \bibinfo {author} {\bibfnamefont
  {I.}~\bibnamefont {Robert-Philip}}, \bibinfo {author} {\bibfnamefont
  {B.}~\bibnamefont {Gil}}, \bibinfo {author} {\bibfnamefont {G.}~\bibnamefont
  {Cassabois}}, \bibinfo {author} {\bibfnamefont {A.}~\bibnamefont {Finco}}, \
  and\ \bibinfo {author} {\bibfnamefont {V.}~\bibnamefont {Jacques}},\ }\href
  {\doibase 10.1103/PhysRevApplied.18.L061002} {\bibfield  {journal} {\bibinfo
  {journal} {Physical Review Applied}\ }\textbf {\bibinfo {volume} {18}},\
  \bibinfo {pages} {L061002} (\bibinfo {year} {2022})},\ \bibinfo {note}
  {publisher: American Physical Society}\BibitemShut {NoStop}%
\bibitem [{\citenamefont {Haykal}\ \emph {et~al.}(2022)\citenamefont {Haykal},
  \citenamefont {Tanos}, \citenamefont {Minotto}, \citenamefont {Durand},
  \citenamefont {Fabre}, \citenamefont {Li}, \citenamefont {Edgar},
  \citenamefont {Iv√°dy}, \citenamefont {Gali}, \citenamefont {Michel},
  \citenamefont {Dr√©au}, \citenamefont {Gil}, \citenamefont {Cassabois},\ and\
  \citenamefont {Jacques}}]{haykal_decoherence_2022}%
  \BibitemOpen
  \bibfield  {author} {\bibinfo {author} {\bibfnamefont {A.}~\bibnamefont
  {Haykal}}, \bibinfo {author} {\bibfnamefont {R.}~\bibnamefont {Tanos}},
  \bibinfo {author} {\bibfnamefont {N.}~\bibnamefont {Minotto}}, \bibinfo
  {author} {\bibfnamefont {A.}~\bibnamefont {Durand}}, \bibinfo {author}
  {\bibfnamefont {F.}~\bibnamefont {Fabre}}, \bibinfo {author} {\bibfnamefont
  {J.}~\bibnamefont {Li}}, \bibinfo {author} {\bibfnamefont {J.~H.}\
  \bibnamefont {Edgar}}, \bibinfo {author} {\bibfnamefont {V.}~\bibnamefont
  {Ivády}}, \bibinfo {author} {\bibfnamefont {A.}~\bibnamefont {Gali}},
  \bibinfo {author} {\bibfnamefont {T.}~\bibnamefont {Michel}}, \bibinfo
  {author} {\bibfnamefont {A.}~\bibnamefont {Dréau}}, \bibinfo {author}
  {\bibfnamefont {B.}~\bibnamefont {Gil}}, \bibinfo {author} {\bibfnamefont
  {G.}~\bibnamefont {Cassabois}}, \ and\ \bibinfo {author} {\bibfnamefont
  {V.}~\bibnamefont {Jacques}},\ }\href {\doibase 10.1038/s41467-022-31743-0}
  {\bibfield  {journal} {\bibinfo  {journal} {Nature Communications}\ }\textbf
  {\bibinfo {volume} {13}},\ \bibinfo {pages} {4347} (\bibinfo {year}
  {2022})},\ \bibinfo {note} {number: 1 Publisher: Nature Publishing
  Group}\BibitemShut {NoStop}%
\bibitem [{\citenamefont {Tran}\ \emph {et~al.}(2016)\citenamefont {Tran},
  \citenamefont {Bray}, \citenamefont {Ford}, \citenamefont {Toth},\ and\
  \citenamefont {Aharonovich}}]{tran_quantum_2016}%
  \BibitemOpen
  \bibfield  {author} {\bibinfo {author} {\bibfnamefont {T.~T.}\ \bibnamefont
  {Tran}}, \bibinfo {author} {\bibfnamefont {K.}~\bibnamefont {Bray}}, \bibinfo
  {author} {\bibfnamefont {M.~J.}\ \bibnamefont {Ford}}, \bibinfo {author}
  {\bibfnamefont {M.}~\bibnamefont {Toth}}, \ and\ \bibinfo {author}
  {\bibfnamefont {I.}~\bibnamefont {Aharonovich}},\ }\href {\doibase
  10.1038/nnano.2015.242} {\bibfield  {journal} {\bibinfo  {journal} {Nature
  Nanotechnology}\ }\textbf {\bibinfo {volume} {11}},\ \bibinfo {pages} {37}
  (\bibinfo {year} {2016})}\BibitemShut {NoStop}%
\bibitem [{\citenamefont {Sajid}\ \emph {et~al.}(2020)\citenamefont {Sajid},
  \citenamefont {Ford},\ and\ \citenamefont
  {Reimers}}]{sajid_single-photon_2020}%
  \BibitemOpen
  \bibfield  {author} {\bibinfo {author} {\bibfnamefont {A.}~\bibnamefont
  {Sajid}}, \bibinfo {author} {\bibfnamefont {M.~J.}\ \bibnamefont {Ford}}, \
  and\ \bibinfo {author} {\bibfnamefont {J.~R.}\ \bibnamefont {Reimers}},\
  }\href {\doibase 10.1088/1361-6633/ab6310} {\bibfield  {journal} {\bibinfo
  {journal} {Reports on Progress in Physics}\ }\textbf {\bibinfo {volume}
  {83}},\ \bibinfo {pages} {044501} (\bibinfo {year} {2020})},\ \bibinfo {note}
  {publisher: IOP Publishing}\BibitemShut {NoStop}%
\bibitem [{\citenamefont {Rayner}\ and\ \citenamefont
  {Brown}(1973)}]{rayner1973crystal}%
  \BibitemOpen
  \bibfield  {author} {\bibinfo {author} {\bibfnamefont {J.}~\bibnamefont
  {Rayner}}\ and\ \bibinfo {author} {\bibfnamefont {G.}~\bibnamefont {Brown}},\
  }\href@noop {} {\bibfield  {journal} {\bibinfo  {journal} {Clays and Clay
  Minerals}\ }\textbf {\bibinfo {volume} {21}},\ \bibinfo {pages} {103}
  (\bibinfo {year} {1973})}\BibitemShut {NoStop}%
\bibitem [{\citenamefont {Ulian}\ \emph {et~al.}(2013)\citenamefont {Ulian},
  \citenamefont {Tosoni},\ and\ \citenamefont
  {Valdr{\`e}}}]{ulian2013comparison}%
  \BibitemOpen
  \bibfield  {author} {\bibinfo {author} {\bibfnamefont {G.}~\bibnamefont
  {Ulian}}, \bibinfo {author} {\bibfnamefont {S.}~\bibnamefont {Tosoni}}, \
  and\ \bibinfo {author} {\bibfnamefont {G.}~\bibnamefont {Valdr{\`e}}},\
  }\href@noop {} {\bibfield  {journal} {\bibinfo  {journal} {The Journal of
  chemical physics}\ }\textbf {\bibinfo {volume} {139}} (\bibinfo {year}
  {2013})}\BibitemShut {NoStop}%
\bibitem [{\citenamefont {Feres}\ \emph {et~al.}()\citenamefont {Feres},
  \citenamefont {Maia}, \citenamefont {Chen}, \citenamefont {Mayer},
  \citenamefont {Obst}, \citenamefont {Hatem}, \citenamefont {Wehmeier},
  \citenamefont {N√∂renberg}, \citenamefont {Queiroz}, \citenamefont
  {Mazzotti}, \citenamefont {Klopf}, \citenamefont {Kehr}, \citenamefont {Eng},
  \citenamefont {Cadore}, \citenamefont {Hillenbrand}, \citenamefont
  {Freitas},\ and\ \citenamefont {Barcelos}}]{feres_two-dimensional_2025}%
  \BibitemOpen
  \bibfield  {author} {\bibinfo {author} {\bibfnamefont {F.~H.}\ \bibnamefont
  {Feres}}, \bibinfo {author} {\bibfnamefont {F.~C.~B.}\ \bibnamefont {Maia}},
  \bibinfo {author} {\bibfnamefont {S.}~\bibnamefont {Chen}}, \bibinfo {author}
  {\bibfnamefont {R.~A.}\ \bibnamefont {Mayer}}, \bibinfo {author}
  {\bibfnamefont {M.}~\bibnamefont {Obst}}, \bibinfo {author} {\bibfnamefont
  {O.}~\bibnamefont {Hatem}}, \bibinfo {author} {\bibfnamefont
  {L.}~\bibnamefont {Wehmeier}}, \bibinfo {author} {\bibfnamefont
  {T.}~\bibnamefont {Nörenberg}}, \bibinfo {author} {\bibfnamefont {M.~S.}\
  \bibnamefont {Queiroz}}, \bibinfo {author} {\bibfnamefont {V.}~\bibnamefont
  {Mazzotti}}, \bibinfo {author} {\bibfnamefont {J.~M.}\ \bibnamefont {Klopf}},
  \bibinfo {author} {\bibfnamefont {S.~C.}\ \bibnamefont {Kehr}}, \bibinfo
  {author} {\bibfnamefont {L.~M.}\ \bibnamefont {Eng}}, \bibinfo {author}
  {\bibfnamefont {A.~R.}\ \bibnamefont {Cadore}}, \bibinfo {author}
  {\bibfnamefont {R.}~\bibnamefont {Hillenbrand}}, \bibinfo {author}
  {\bibfnamefont {R.~O.}\ \bibnamefont {Freitas}}, \ and\ \bibinfo {author}
  {\bibfnamefont {I.~D.}\ \bibnamefont {Barcelos}},\ }\href {\doibase
  10.48550/arXiv.2501.17340} {\enquote {\bibinfo {title} {Two-dimensional talc
  as a natural hyperbolic material},}\ }\Eprint
  {http://arxiv.org/abs/2501.17340 [cond-mat]} {2501.17340 [cond-mat]}
  \BibitemShut {NoStop}%
\bibitem [{\citenamefont {Robinson}()}]{robinson_measurement_2004}%
  \BibitemOpen
  \bibfield  {author} {\bibinfo {author} {\bibfnamefont {D.~A.}\ \bibnamefont
  {Robinson}},\ }\href {\doibase 10.2136/vzj2004.0705} {\ \textbf {\bibinfo
  {volume} {3}},\ \bibinfo {pages} {705}} \BibitemShut
  {NoStop}%
\bibitem [{\citenamefont {Freysoldt}\ \emph {et~al.}(2014)\citenamefont
  {Freysoldt}, \citenamefont {Grabowski}, \citenamefont {Hickel}, \citenamefont
  {Neugebauer}, \citenamefont {Kresse}, \citenamefont {Janotti},\ and\
  \citenamefont {Van~de Walle}}]{freysoldt_first-principles_2014}%
  \BibitemOpen
  \bibfield  {author} {\bibinfo {author} {\bibfnamefont {C.}~\bibnamefont
  {Freysoldt}}, \bibinfo {author} {\bibfnamefont {B.}~\bibnamefont
  {Grabowski}}, \bibinfo {author} {\bibfnamefont {T.}~\bibnamefont {Hickel}},
  \bibinfo {author} {\bibfnamefont {J.}~\bibnamefont {Neugebauer}}, \bibinfo
  {author} {\bibfnamefont {G.}~\bibnamefont {Kresse}}, \bibinfo {author}
  {\bibfnamefont {A.}~\bibnamefont {Janotti}}, \ and\ \bibinfo {author}
  {\bibfnamefont {C.~G.}\ \bibnamefont {Van~de Walle}},\ }\href {\doibase
  10.1103/RevModPhys.86.253} {\bibfield  {journal} {\bibinfo  {journal}
  {Reviews of Modern Physics}\ }\textbf {\bibinfo {volume} {86}},\ \bibinfo
  {pages} {253} (\bibinfo {year} {2014})},\ \bibinfo {note} {publisher:
  American Physical Society}\BibitemShut {NoStop}%
\bibitem [{\citenamefont {Zhang}\ \emph {et~al.}(2023)\citenamefont {Zhang},
  \citenamefont {Yan}, \citenamefont {Qiu}, \citenamefont {Zhang},
  \citenamefont {Shen}, \citenamefont {Wei},\ and\ \citenamefont
  {Deng}}]{zhang_correcting_2023}%
  \BibitemOpen
  \bibfield  {author} {\bibinfo {author} {\bibfnamefont {C.}~\bibnamefont
  {Zhang}}, \bibinfo {author} {\bibfnamefont {L.}~\bibnamefont {Yan}}, \bibinfo
  {author} {\bibfnamefont {C.}~\bibnamefont {Qiu}}, \bibinfo {author}
  {\bibfnamefont {C.-X.}\ \bibnamefont {Zhang}}, \bibinfo {author}
  {\bibfnamefont {T.}~\bibnamefont {Shen}}, \bibinfo {author} {\bibfnamefont
  {S.-H.}\ \bibnamefont {Wei}}, \ and\ \bibinfo {author} {\bibfnamefont
  {H.-X.}\ \bibnamefont {Deng}},\ }\href {\doibase 10.1103/PhysRevB.108.245305}
  {\bibfield  {journal} {\bibinfo  {journal} {Physical Review B}\ }\textbf
  {\bibinfo {volume} {108}},\ \bibinfo {pages} {245305} (\bibinfo {year}
  {2023})}\BibitemShut {NoStop}%
\bibitem [{\citenamefont {Weston}\ \emph {et~al.}(2018)\citenamefont {Weston},
  \citenamefont {Wickramaratne}, \citenamefont {Mackoit}, \citenamefont
  {Alkauskas},\ and\ \citenamefont {Van~de Walle}}]{weston_native_2018}%
  \BibitemOpen
  \bibfield  {author} {\bibinfo {author} {\bibfnamefont {L.}~\bibnamefont
  {Weston}}, \bibinfo {author} {\bibfnamefont {D.}~\bibnamefont
  {Wickramaratne}}, \bibinfo {author} {\bibfnamefont {M.}~\bibnamefont
  {Mackoit}}, \bibinfo {author} {\bibfnamefont {A.}~\bibnamefont {Alkauskas}},
  \ and\ \bibinfo {author} {\bibfnamefont {C.~G.}\ \bibnamefont {Van~de
  Walle}},\ }\href {\doibase 10.1103/PhysRevB.97.214104} {\bibfield  {journal}
  {\bibinfo  {journal} {Physical Review B}\ }\textbf {\bibinfo {volume} {97}},\
  \bibinfo {pages} {214104} (\bibinfo {year} {2018})},\ \bibinfo {note}
  {publisher: American Physical Society}\BibitemShut {NoStop}%
\bibitem [{\citenamefont {Thiering}\ and\ \citenamefont
  {Gali}(2016)}]{thiering_characterization_2016}%
  \BibitemOpen
  \bibfield  {author} {\bibinfo {author} {\bibfnamefont {G.}~\bibnamefont
  {Thiering}}\ and\ \bibinfo {author} {\bibfnamefont {A.}~\bibnamefont
  {Gali}},\ }\href {\doibase 10.1103/PhysRevB.94.125202} {\bibfield  {journal}
  {\bibinfo  {journal} {Physical Review B}\ }\textbf {\bibinfo {volume} {94}},\
  \bibinfo {pages} {125202} (\bibinfo {year} {2016})},\ \bibinfo {note}
  {publisher: American Physical Society}\BibitemShut {NoStop}%
\bibitem [{\citenamefont {Benedek}\ \emph {et~al.}()\citenamefont {Benedek},
  \citenamefont {Ganyecz}, \citenamefont {Pershin}, \citenamefont {Iv√°dy},\
  and\ \citenamefont {Barcza}}]{benedek_accurate_2024}%
  \BibitemOpen
  \bibfield  {author} {\bibinfo {author} {\bibfnamefont {Z.}~\bibnamefont
  {Benedek}}, \bibinfo {author} {\bibfnamefont {Á.}~\bibnamefont {Ganyecz}},
  \bibinfo {author} {\bibfnamefont {A.}~\bibnamefont {Pershin}}, \bibinfo
  {author} {\bibfnamefont {V.}~\bibnamefont {Ivády}}, \ and\ \bibinfo {author}
  {\bibfnamefont {G.}~\bibnamefont {Barcza}},\ }\href {\doibase
  10.48550/arXiv.2406.05092} {\enquote {\bibinfo {title} {Accurate and
  convergent energetics of color centers by wavefunction theory},}\ }\Eprint
  {http://arxiv.org/abs/2406.05092 [cond-mat]} {2406.05092 [cond-mat]}
  \BibitemShut {NoStop}%
\bibitem [{\citenamefont {Ivády}\ \emph {et~al.}(2013)\citenamefont {Ivády},
  \citenamefont {Abrikosov}, \citenamefont {Janzén},\ and\ \citenamefont
  {Gali}}]{ivady_role_2013}%
  \BibitemOpen
  \bibfield  {author} {\bibinfo {author} {\bibfnamefont {V.}~\bibnamefont
  {Ivády}}, \bibinfo {author} {\bibfnamefont {I.~A.}\ \bibnamefont
  {Abrikosov}}, \bibinfo {author} {\bibfnamefont {E.}~\bibnamefont {Janzén}},
  \ and\ \bibinfo {author} {\bibfnamefont {A.}~\bibnamefont {Gali}},\ }\href
  {\doibase 10.1103/PhysRevB.87.205201} {\bibfield  {journal} {\bibinfo
  {journal} {Physical Review B}\ }\textbf {\bibinfo {volume} {87}},\ \bibinfo
  {pages} {205201} (\bibinfo {year} {2013})},\ \bibinfo {note} {publisher:
  American Physical Society}\BibitemShut {NoStop}%
\bibitem [{\citenamefont {Ivády}\ \emph {et~al.}(2014)\citenamefont {Ivády},
  \citenamefont {Armiento}, \citenamefont {Szász}, \citenamefont {Janzén},
  \citenamefont {Gali},\ and\ \citenamefont
  {Abrikosov}}]{ivady_theoretical_2014}%
  \BibitemOpen
  \bibfield  {author} {\bibinfo {author} {\bibfnamefont {V.}~\bibnamefont
  {Ivády}}, \bibinfo {author} {\bibfnamefont {R.}~\bibnamefont {Armiento}},
  \bibinfo {author} {\bibfnamefont {K.}~\bibnamefont {Szász}}, \bibinfo
  {author} {\bibfnamefont {E.}~\bibnamefont {Janzén}}, \bibinfo {author}
  {\bibfnamefont {A.}~\bibnamefont {Gali}}, \ and\ \bibinfo {author}
  {\bibfnamefont {I.~A.}\ \bibnamefont {Abrikosov}},\ }\href {\doibase
  10.1103/PhysRevB.90.035146} {\bibfield  {journal} {\bibinfo  {journal}
  {Physical Review B}\ }\textbf {\bibinfo {volume} {90}},\ \bibinfo {pages}
  {035146} (\bibinfo {year} {2014})},\ \bibinfo {note} {publisher: American
  Physical Society}\BibitemShut {NoStop}%
\bibitem [{\citenamefont {Tetienne}(2021)}]{tetienne_quantum_2021}%
  \BibitemOpen
  \bibfield  {author} {\bibinfo {author} {\bibfnamefont {J.-P.}\ \bibnamefont
  {Tetienne}},\ }\href {\doibase 10.1038/s41567-021-01338-5} {\bibfield
  {journal} {\bibinfo  {journal} {Nature Physics}\ ,\ \bibinfo {pages} {1}}
  (\bibinfo {year} {2021})},\ \bibinfo {note} {bandiera\_abtest: a Cg\_type:
  Nature Research Journals Primary\_atype: News \& Views Publisher: Nature
  Publishing Group Subject\_term: Qubits;Two-dimensional materials
  Subject\_term\_id: qubits;two-dimensional-materials}\BibitemShut {NoStop}%
\bibitem [{\citenamefont {Kresse}\ and\ \citenamefont {Hafner}(1994)}]{VASP}%
  \BibitemOpen
  \bibfield  {author} {\bibinfo {author} {\bibfnamefont {G.}~\bibnamefont
  {Kresse}}\ and\ \bibinfo {author} {\bibfnamefont {J.}~\bibnamefont
  {Hafner}},\ }\href {\doibase 10.1103/PhysRevB.49.14251} {\bibfield  {journal}
  {\bibinfo  {journal} {Phys. Rev. B}\ }\textbf {\bibinfo {volume} {49}},\
  \bibinfo {pages} {14251} (\bibinfo {year} {1994})}\BibitemShut {NoStop}%
\bibitem [{\citenamefont {Kresse}\ and\ \citenamefont
  {Furthm\"uller}(1996)}]{VASP2}%
  \BibitemOpen
  \bibfield  {author} {\bibinfo {author} {\bibfnamefont {G.}~\bibnamefont
  {Kresse}}\ and\ \bibinfo {author} {\bibfnamefont {J.}~\bibnamefont
  {Furthm\"uller}},\ }\href {\doibase 10.1103/PhysRevB.54.11169} {\bibfield
  {journal} {\bibinfo  {journal} {Phys. Rev. B}\ }\textbf {\bibinfo {volume}
  {54}},\ \bibinfo {pages} {11169} (\bibinfo {year} {1996})}\BibitemShut
  {NoStop}%
\bibitem [{\citenamefont {Bl\"ochl}(1994)}]{PAW}%
  \BibitemOpen
  \bibfield  {author} {\bibinfo {author} {\bibfnamefont {P.~E.}\ \bibnamefont
  {Bl\"ochl}},\ }\href {\doibase 10.1103/PhysRevB.50.17953} {\bibfield
  {journal} {\bibinfo  {journal} {Phys. Rev. B}\ }\textbf {\bibinfo {volume}
  {50}},\ \bibinfo {pages} {17953} (\bibinfo {year} {1994})}\BibitemShut
  {NoStop}%
\bibitem [{\citenamefont {Heyd}\ \emph {et~al.}(2003)\citenamefont {Heyd},
  \citenamefont {Scuseria},\ and\ \citenamefont
  {Ernzerhof}}]{heyd_hybrid_2003}%
  \BibitemOpen
  \bibfield  {author} {\bibinfo {author} {\bibfnamefont {J.}~\bibnamefont
  {Heyd}}, \bibinfo {author} {\bibfnamefont {G.~E.}\ \bibnamefont {Scuseria}},
  \ and\ \bibinfo {author} {\bibfnamefont {M.}~\bibnamefont {Ernzerhof}},\
  }\href {\doibase 10.1063/1.1564060} {\ \textbf {\bibinfo {volume} {118}},\
  \bibinfo {pages} {8207} (\bibinfo {year} {2003})}\BibitemShut {NoStop}%
\bibitem [{\citenamefont {Heyd}\ \emph {et~al.}(2006)\citenamefont {Heyd},
  \citenamefont {Scuseria},\ and\ \citenamefont
  {Ernzerhof}}]{heyd_erratum:_2006}%
  \BibitemOpen
  \bibfield  {author} {\bibinfo {author} {\bibfnamefont {J.}~\bibnamefont
  {Heyd}}, \bibinfo {author} {\bibfnamefont {G.~E.}\ \bibnamefont {Scuseria}},
  \ and\ \bibinfo {author} {\bibfnamefont {M.}~\bibnamefont {Ernzerhof}},\
  }\href {\doibase 10.1063/1.2204597} {\ \textbf {\bibinfo {volume} {124}},\
  \bibinfo {pages} {219906} (\bibinfo {year} {2006})}\BibitemShut {NoStop}%
\bibitem [{\citenamefont {Henderson}\ \emph {et~al.}(2011)\citenamefont
  {Henderson}, \citenamefont {Paier},\ and\ \citenamefont
  {Scuseria}}]{henderson_accurate_2011}%
  \BibitemOpen
  \bibfield  {author} {\bibinfo {author} {\bibfnamefont {T.~M.}\ \bibnamefont
  {Henderson}}, \bibinfo {author} {\bibfnamefont {J.}~\bibnamefont {Paier}}, \
  and\ \bibinfo {author} {\bibfnamefont {G.~E.}\ \bibnamefont {Scuseria}},\
  }\href {\doibase 10.1002/pssb.201046303} {\bibfield  {journal} {\bibinfo
  {journal} {physica status solidi (b)}\ }\textbf {\bibinfo {volume} {248}},\
  \bibinfo {pages} {767} (\bibinfo {year} {2011})} \BibitemShut
  {NoStop}%
\bibitem [{\citenamefont {Grimme}\ \emph {et~al.}(2010)\citenamefont {Grimme},
  \citenamefont {Antony}, \citenamefont {Ehrlich},\ and\ \citenamefont
  {Krieg}}]{grimme_consistent_2010}%
  \BibitemOpen
  \bibfield  {author} {\bibinfo {author} {\bibfnamefont {S.}~\bibnamefont
  {Grimme}}, \bibinfo {author} {\bibfnamefont {J.}~\bibnamefont {Antony}},
  \bibinfo {author} {\bibfnamefont {S.}~\bibnamefont {Ehrlich}}, \ and\
  \bibinfo {author} {\bibfnamefont {H.}~\bibnamefont {Krieg}},\ }\href
  {\doibase 10.1063/1.3382344} {\bibfield  {journal} {\bibinfo  {journal} {The
  Journal of Chemical Physics}\ }\textbf {\bibinfo {volume} {132}},\ \bibinfo
  {pages} {154104} (\bibinfo {year} {2010})}\BibitemShut {NoStop}%
\bibitem [{\citenamefont {Gali}\ \emph {et~al.}(2009)\citenamefont {Gali},
  \citenamefont {Janz√©n}, \citenamefont {De√°k}, \citenamefont {Kresse},\ and\
  \citenamefont {Kaxiras}}]{gali_theory_2009}%
  \BibitemOpen
  \bibfield  {author} {\bibinfo {author} {\bibfnamefont {A.}~\bibnamefont
  {Gali}}, \bibinfo {author} {\bibfnamefont {E.}~\bibnamefont {Janzén}},
  \bibinfo {author} {\bibfnamefont {P.}~\bibnamefont {Deák}}, \bibinfo
  {author} {\bibfnamefont {G.}~\bibnamefont {Kresse}}, \ and\ \bibinfo {author}
  {\bibfnamefont {E.}~\bibnamefont {Kaxiras}},\ }\href@noop {} {\bibfield
  {journal} {\bibinfo  {journal} {Phys. Rev. Lett.}\ }\textbf {\bibinfo
  {volume} {103}},\ \bibinfo {pages} {186404} (\bibinfo {year}
  {2009})}\BibitemShut {NoStop}%
\bibitem [{\citenamefont {Szász}\ \emph {et~al.}(2013)\citenamefont {Szász},
  \citenamefont {Hornos}, \citenamefont {Marsman},\ and\ \citenamefont
  {Gali}}]{szasz_hyperfine_2013}%
  \BibitemOpen
  \bibfield  {author} {\bibinfo {author} {\bibfnamefont {K.}~\bibnamefont
  {Szász}}, \bibinfo {author} {\bibfnamefont {T.}~\bibnamefont {Hornos}},
  \bibinfo {author} {\bibfnamefont {M.}~\bibnamefont {Marsman}}, \ and\
  \bibinfo {author} {\bibfnamefont {A.}~\bibnamefont {Gali}},\ }\href {\doibase
  10.1103/PhysRevB.88.075202} {\bibfield  {journal} {\bibinfo  {journal}
  {Physical Review B}\ }\textbf {\bibinfo {volume} {88}},\ \bibinfo {pages}
  {075202} (\bibinfo {year} {2013})}\BibitemShut {NoStop}%
\bibitem [{\citenamefont {Takács}\ and\ \citenamefont
  {Ivády}()}]{takacs_accurate_2024}%
  \BibitemOpen
  \bibfield  {author} {\bibinfo {author} {\bibfnamefont {I.}~\bibnamefont
  {Takács}}\ and\ \bibinfo {author} {\bibfnamefont {V.}~\bibnamefont
  {Ivády}},\ }\href {\doibase 10.1038/s42005-024-01668-9} {\ \textbf {\bibinfo
  {volume} {7}},\ \bibinfo {pages} {1}},\ \bibinfo {note} {publisher: Nature
  Publishing Group}\BibitemShut {NoStop}%
\bibitem [{\citenamefont {Dancoff}(1950)}]{dancoffNonAdiabaticMesonTheory1950}%
  \BibitemOpen
  \bibfield  {author} {\bibinfo {author} {\bibfnamefont {S.~M.}\ \bibnamefont
  {Dancoff}},\ }\href {\doibase 10.1103/PhysRev.78.382} {\bibfield  {journal}
  {\bibinfo  {journal} {Physical Review}\ }\textbf {\bibinfo {volume} {78}},\
  \bibinfo {pages} {382} (\bibinfo {year} {1950})}\BibitemShut {NoStop}%
\bibitem [{\citenamefont {Tamm}\ \emph {et~al.}(2011)\citenamefont {Tamm},
  \citenamefont {Peierls}, \citenamefont {Bolotovskii},\ and\ \citenamefont
  {Frenkel}}]{tamm2011a}%
  \BibitemOpen
  \bibfield  {author} {\bibinfo {author} {\bibfnamefont {I.~E.}\ \bibnamefont
  {Tamm}}, \bibinfo {author} {\bibfnamefont {R.}~\bibnamefont {Peierls}},
  \bibinfo {author} {\bibfnamefont {B.~M.}\ \bibnamefont {Bolotovskii}}, \ and\
  \bibinfo {author} {\bibfnamefont {V.~Y.}\ \bibnamefont {Frenkel}},\ }in\
  \href {https://books.google.se/books?id=Xg2EMAEACAAJ} {\emph {\bibinfo
  {booktitle} {Selected {{Papers}}}}}\ (\bibinfo  {publisher} {Springer Berlin
  Heidelberg},\ \bibinfo {year} {2011})\ pp.\ \bibinfo {pages}
  {157--174}\BibitemShut {NoStop}%
\end{thebibliography}

%

\end{document}